\documentclass[12pt,oneside,subeqn]{article}
\usepackage{amssymb,amsthm}
\usepackage{amsfonts}
\usepackage{verbatim}
\usepackage{psfrag,graphicx}
\usepackage{amsmath,natbib}
\usepackage{setspace}
\usepackage[left = 1.2in, top = 4in]{geometry}
\usepackage{pstricks}
\usepackage[pagewise,mathlines]{lineno}
\usepackage{psfrag,graphicx}\graphicspath{{./figures/}}
\usepackage{amsmath,natbib}
\usepackage{setspace}
\usepackage{xspace}
\usepackage{subfigure}
\usepackage{geometry,setspace}
\usepackage[page]{appendix}
\usepackage{color}
\newtheorem{thm}{Theorem}

\newtheorem{alg}{Algorithm}
\newtheorem{lemma}{Lemma}
\newtheorem{corollary}{Corollary}

\newcommand{\ttheta}{\tilde \theta}
\newcommand{\apf}{auxiliary particle filter\xspace}

\newcommand{\MCMC}{Markov chain Monte Carlo\xspace}
\newcommand{\MH}{Metropolis-Hastings\xspace}

\newcommand{\Uhat}{\widehat{U}}

\newcommand{\phat}{\widehat{p}}

\newcommand{\aimh}{adaptive independent Metropolis Hastings\xspace}
\newcommand{\imh}{independent Metropolis Hastings\xspace}

\newcommand{\mn}{mixture of normals\xspace}
\newcommand{\arwm}{adaptive random walk Metropolis\xspace}
\newcommand{\xtilde}{\widetilde {x}}

\newcommand{\thetat}{\widetilde \theta}

\newcommand{\pf}{particle filter\xspace}
\newcommand{\ra}{\rightarrow}

\newcommand{\ssm}{state space model\xspace}
\newcommand{\gssm}{Gaussian state space model\xspace}
\newcommand{\tsum}{\sum}
\newcommand{\tint}{\int}
\newcommand{\Bin}{\text{\rm Bin}}
\newcommand{\KF}{Kalman filter}
\newcommand{\papf}{partially adapted \pf}
\newcommand{\fapf}{fully adapted \pf}
\newcommand{\spf}{standard \pf}
\newcommand{\Var}{\text{Var}}
\newcommand{\sigsq}{\sigma^2}
\newcommand{\xbreve}{\breve x}
\newcommand{\Sigmabreve}{\breve \Sigma}
\newcommand{\half}{\frac12}
\newcommand{\clt}{cental limit theorem,\xspace}

\DeclareMathOperator*{\argmax}{arg\,max}
\newcommand{\logit}{\text{logit}}

\newcommand{\up}[1]{\raisebox{0.25cm}[0pt]{#1}}

\newcommand{\blind}{0}

\def\spacingset#1{\renewcommand{\baselinestretch}%
{#1}\small\normalsize} \spacingset{1}

\if0\blind
{

\title{\textbf{Auxiliary particle filtering within adaptive \MH{} sampling}}

\author{Michael K. Pitt\\
        {\small Economics Department}\\
        {\small University of Warwick}\\
        {\small m.pitt@warwick.ac.uk}
\and Ralph S. Silva\\
        {\small School of Economics}\\
        {\small University of New South Wales}\\
        {\small r.silva@unsw.edu.au}
\and Paolo Giordani\\
        {\small Research Department}\\
        {\small Sveriges Riksbank}\\
        {\small paolo.giordani@riksbank.se}\\[0.5cm]
\and Robert Kohn\footnote{Corresponding author.}\\
        {\small School of Economics}\\
        {\small University of New South Wales}\\
        {\small r.kohn@unsw.edu.au}
}
\fi

\if1\blind
{
 \bigskip
  \bigskip
  \bigskip
\title{\textbf{Particle filtering within adaptive \MH sampling}}
\medskip
}
\fi

\date{\vspace{0.5cm}{\small May 15 2010}}
\begin{document}
\maketitle
\thispagestyle{empty}
\begin{abstract}
Our article deals with Bayesian inference for a general state space model with
the simulated likelihood computed by the \pf. We show empirically that
the partially or \fapf{s} can be much more efficient than the
\spf, especially when the signal to noise ratio is high.
This is especially important because using the \pf within \MCMC sampling is $O(T^2)$,
where $T$ is the sample size. We also show that an \aimh proposal for the unknown parameters based on a \mn  can be much more efficient than the usual optimal random walk
methods because the simulated likelihood is not continuous in the parameters and
the cost of constructing a good adaptive proposal is negligible compared
to the cost of evaluating the simulated likelihood.
Independent \MH proposals are also attractive because they are easy to
run in parallel on multiple processors. The article also shows that the proposed \aimh
sampler converges to the posterior distribution.
We also show that the marginal
likelihood of any state space model can be
obtained in an efficient and unbiased manner by using the \pf making model
comparison straightforward. Obtaining the marginal likelihood is often difficult
using other methods. Finally, we prove
that the simulated likelihood obtained by the \apf is unbiased.
This result is fundamental to using the \pf for \MCMC sampling
and is first obtained in a more abstract and difficult setting by
\cite{delmoral:2004}. However, our proof is direct and will make the result accessible
to readers.

\end{abstract}

\spacingset{1.0}
\section{Introduction}\label{section:introduction}
Our article builds on the work of \citet{andrieu:doucet:holenstein:2010}
and develops general simulation methods that make Bayesian inference
for time series \ssm{s} feasible and efficient. Our first contribution is to
show empirically that partially or fully adapted \apf{s} using the simulated
likelihood defined in \cite{pitt:2002} can be much more efficient
statistically than the standard \spf of
\cite{gordon:salmond:smith:1993}, especially when the signal to noise ratio is high,
because they reduce the noise in the simulated likelihood.
It is very important to carry out the \pf as efficiently as possible,
because \pf{ing} within \MCMC sampling
is $O(T^2)$, where $T$ is the sample size as explained in
Section~\ref{SS: general asir method}.

Adaptive sampling methods are simulation methods for carrying out Bayesian inference
that use previous iterates of the simulation  to form proposal distributions,
that is, the adaptive
samplers learn about some aspects of the posterior distribution from previous iterates.
See for example \cite{haario:saksman:tamminen:2001},
\cite{atchade:rosenthal:2005} and  \cite{roberts:rosenthal:2008}
who consider \arwm proposals and \citet{giordani:kohn:2010}
who base their proposal on a \mn{}.

The second contribution of the article is to show that when working with a
simulated likelihood  it is worthwhile constructing \aimh proposals
that provide good approximations to the posterior density. The first reason for this claim is that the simulated likelihood is not continuous in the unknown parameters.
This means that standard methods for constructing proposals such as Laplace approximations based on analytic or numerical derivatives are usually infeasible. It also means that
the usual optimal random walk methods do not perform as well as expected
as the probability of acceptance does not tend to 1 as a proposed move becomes
more local moves or even if the parameter does not change at all.
The second reason is that the cost of constructing a good adaptive proposal
is often negligible compared to the cost of running the \pf to obtain the simulated likelihood. Third, an adaptive sampling scheme that consists entirely or mostly of independent \MH steps is attractive because a large part of the computation can be run
in parallel thus substantially reducing computing time.

Our article uses the \aimh sampler of \citet{giordani:kohn:2010}
which approximates the posterior density
by a \mn. We show that this proposal density can be much more efficient
than the \arwm proposal of \cite{roberts:rosenthal:2008} for the reasons just outlined.
We also show that this adaptive sampler converges to the correct posterior distribution.
We note, however,  that in our experience, the \arwm algorithm
of \cite{roberts:rosenthal:2008} is important because it converges reliably
for a diverse set of problems and  provides a good way to initialize
other more efficient adaptive sampling schemes.

The third contribution of our article is to show that the marginal likelihood
of any  \ssm  can be estimated  in an efficient and unbiased manner
by combining \pf{ing} with bridge or importance sampling.
This makes it straightforward to compare the marginal likelihoods of
two or more models each of which  can be expressed in state space form.

The final contribution of the article is to show that the simulated likelihood
obtained the \apf is unbiased. This result is obtained in an abstract setting
in Proposition 7.4.1
in Section 7.4.2 in \cite{delmoral:2004}.
\citet{andrieu:doucet:holenstein:2010} show that the unbiasedness of the
simulated likelihood allows Bayesian inference using \MCMC simulation.
This is because the simulated likelihood can viewed as the density of the
observations conditional on the unknown parameters and a set of auxiliary latent variables.
We believe that our derivation of unbiasedness
is more direct and accessible than that of \cite{delmoral:2004}.

Computational algorithms for state space models such as the Kalman filter and \pf
are useful because many time series models can be expressed in state space form. Computational methods for Bayesian inference for
\gssm are well developed \citep[see][]{cappe:moulines:ryden:2005}
and there is a literature now on Bayesian computational methods
for non-Gaussian \ssm{s}. \MCMC computational methods based on the \pf have
the potential to greatly increase the number and complexity of time series models
amenable to Bayesian analysis. An early of the \pf within an \MCMC framework is by
\cite{fernández2007estimating} who applied it to macroeconomic models as an approximate approach
for obtaining the posterior distribution of the parameters.

Particle filtering (also known as sequential Monte Carlo)
was proposed by \cite{gordon:salmond:smith:1993}
for online filtering and prediction of nonlinear or non-Gaussian state space models.
The auxiliary particle filter method was introduced by \cite{pitt:shephard:1999}  to improve the performance of the standard particle filter when the observation equation is informative relative to the state equations, that is when the signal to noise ratio is moderate to high.  There is an extensive literature on online filtering  using the particle filter, see for example \cite{kitagawa:1996}, \cite{liu:chen:1998}, \cite{doucet:godsill:andrieu:2000}, \cite{doucet:freitas:gordon:2001}, \cite{andrieu:doucet:2002}, \cite{fearnhead:clifford:2003} and \cite{delmoral:doucet:jasra:2006}. Our article considers only the standard particle filter of \cite{gordon:salmond:smith:1993}  and the fully and \papf{s} proposed by \cite{pitt:shephard:1999}.

The literature on using the particle filter to learn about model parameters is more limited.
\if0\blind{ \cite{pitt:2002}}\fi{} \if1\blind { \cite{XXX}}\fi{}
proposes the smooth \pf{} to estimate the parameters of a state space using maximum likelihood. \cite{storvik:2002} and \cite{polson:stroud:muller:2008} consider online parameter learning when sufficient statistics are available.  \citet{andrieu:doucet:holenstein:2010} provide a
framework for off-line parameter learning using the \pf.
\citet{flury:shephard:2008} give an insightful discussion of the results of \citet{andrieu:doucet:holenstein:2010}
and use single parameter random walk proposals for  off-line Bayesian inference.

Our article is an updated version of  \cite{silva:giordani:kohn:pitt:2009}, which contains some extra 
examples.

\section{State space models}\label{section:PF}
Consider a state space model with observation equation $p(y_t|x_t;\theta)$ and state transition equation
$p(x_t|x_{t-1};\theta)$, where $y_t$ and $x_t$ are the observation and the state at time $t$ and $\theta$
is a vector of unknown parameters. The distribution of the initial state is
$p(x_0|\theta)$.  {See \citet{cappe:moulines:ryden:2005} for a modern treatment of general state space models}.
The filtering equations for  the state space model (for $t \geq 1$) are
\citep[pp.~506-507]{west:harrison:1997}
\begin{subequations}
\begin{align}
p(x_t|y_{1:t-1};\theta) & =\int p(x_t|x_{t-1};\theta)p(x_{t-1}|y_{1:t-1};\theta)dx_{t-1},  \label{eq: update1}\\
p(x_t|y_{1:t};\theta)&= \dfrac{p(y_t|x_{t};\theta)p(x_{t}|y_{1:t-1};\theta)}{p(y_t|y_{1:t-1};\theta)}, \label{eq: update2}\\
p(y_t|y_{1:t-1};\theta) &= \int p(y_t|x_{t};\theta)p(x_{t}|y_{1:t-1};\theta)dx_t \label{eq: one term likelihood}.
\end{align}
\end{subequations}
where $y_{1:t} = \{y_1, \dots, y_t\}$.  Equations~\eqref{eq: update1}--\eqref{eq: one term likelihood}  allow
(in principle) for filtering for a given $\theta$ and for evaluating the likelihood of the observations $y = y_{1:T}$,
\begin{align}\label{eq: likelihood}
p(y|\theta) & = p(y_1|\theta) \prod_{t=1}^{T-1} p(y_{t+1}|y_{1:t};\theta)\ .
\end{align}
If the likelihood $p(y|\theta)$ can be computed, maximum likelihood and MCMC methods can be used to carry out inference
on the parameters $\theta$, with the states integrated out.
When both the observation and state transition equations are linear and Gaussian the likelihood can be
evaluated analytically using the Kalman filter
\citep[pp.~141-143]{cappe:moulines:ryden:2005}. More general state space models can also be estimated
by MCMC methods if auxiliary latent variables are introduced, e.g. \citet{kim:shephard:chib:1998} and \citet{fruhwirhschnatter:wagner:2006} and/or the states are sampled in blocks
as in \citet{shephard:pitt:1997}. See section 6.3 of \citet{cappe:moulines:ryden:2005} for a review of Markov chain Monte Carlo methods applied to general state space models.

In general, however, the integrals in equations
\eqref{eq: update1}--\eqref{eq: one term likelihood} are computationally intractable and the \spf algorithm (SIR)
was  proposed by \cite{gordon:salmond:smith:1993} as a method for approximating them with
the approximation becoming exact as the number of particles tends to infinity.
\cite{pitt:shephard:1999} propose the \apf method (ASIR) which is more efficient than
\spf  when the observation density is informative relative to the transition density.
The general  \apf is described in Section~\ref{SS: general asir method}.

\subsection{General ASIR \ method}\label{SS: general asir method}
The general auxiliary SIR (ASIR) filter of {\normalsize \cite%
{pitt:shephard:1999} may be thought of as a generalisation of the SIR\
method of \cite{gordon:salmond:smith:1993}. We  therefore focus on
this, more general, approach. To simplify notation in this section, we
omit to show dependence on the unknown parameter vector $\theta $. The
following algorithm describes the one time step ASIR update and is initialized
with samples $x_{0}^{k}\sim p(x_{0})$ with mass $1/M$
for $k=1,...,M$.

\bigskip

\begin{alg} \label{alg:asir}{\sl Given samples }$x_{t}^{k}\sim p(x_{t}|y_{1:t})${\sl \
with mass }$\pi _{t}^{k}$ {\sl for }$k=1,...,M.$

{\sl For }${\it t=0,..,T-1:}$

\begin{enumerate}
\item {\sl For }$k=1:M,${\sl \ compute }$\omega
_{t|t+1}^{k}=g(y_{t+1}|x_{t}^{k})\pi _{t}^{k},$ \ \ \ \ $\pi _{t|t+1}^{k}=%
\frac{\omega _{t|t+1}^{k}}{\sum_{i=1}^{M}\omega _{t|t+1}^{i}}.$

\item {\sl For }$k=1:M,${\sl \ sample }$\widetilde{x}_{t}^{k}\sim
\sum_{i=1}^{M}\pi _{t|t+1}^{i}\delta (x_{t}-x_{t}^{i}).$

\item {\sl For }$k=1:M,${\sl \ sample }$x_{t+1}^{k}\sim g(x_{t+1}|\widetilde{%
x}_{t}^{k};y_{t+1}).$

\item {\normalsize \textsl{For }$k=1:M,$\textsl{\ compute}%
\begin{equation*}
\omega _{t+1}^{k}=\frac{p(y_{t+1}|x_{t+1}^{k})p(x_{t+1}^{k}|\widetilde{x}%
_{t}^{k})}{g(y_{t+1}|\widetilde{x}_{t}^{k})g(x_{t+1}^{k}|\widetilde{x}%
_{t}^{k};y_{t+1})},\text{ \ \ \ \ \ \ }\pi _{t+1}^{k}=\frac{\omega _{t+1}^{k}%
}{\sum_{i=1}^{M}\omega _{t+1}^{i}}.
\end{equation*}%
}
\end{enumerate}
\end{alg}

Note that in Step~2,
$\delta (x-a)$ is the delta function with unit mass at $x = a$. In addition,
in Step 2, multinomial sampling may be
employed but stratified sampling is generally to be preferred and is
employed throughout, see \cite{kitagawa:1996},
\cite{carpenter:clifford:fearnhead:1999} and \cite{pitt:shephard:2001}.

Note that the true joint density may be written as,%
\begin{align*}
p(y_{t+1}|x_{t+1})p(x_{t+1}|x_{t})=p(y_{t+1}|x_{t})p(x_{t+1}|x_{t};y_{t+1})%
\text{,}
\end{align*}%
where
\begin{align*}
p(y_{t+1}|x_{t}) &=\tint p(y_{t+1}|x_{t+1})p(x_{t+1}|x_{t})dx_{t+1}, \\
p(x_{t+1}|x_{t};y_{t+1})
&=p(y_{t+1}|x_{t+1})p(x_{t+1}|x_{t})/p(y_{t+1}|x_{t}).
\end{align*}%
Typically this fully adapted form is unavailable but when it is the
approximating joint density may be chosen to be the true joint. That is,
\begin{align*}
g(y_{t+1}|x_{t})g(x_{t+1}|x_{t};y_{t+1}) & =p(y_{t+1}|x_{t})p(x_{t+1}|x_{t};y_{t+1}).
\end{align*}%
In this case Step 4 becomes redundant as $\omega _{t+1}^{k}=1
$, ( $\pi _{t+1}^{k}=1/M$) and the method reduces to what \cite%
{pitt:shephard:2001} call the fully adapted algorithm. The fully adapted
method is the most efficient in estimating the likelihood and is generally
the optimal filter a single time step ahead.

The  SIR method of \cite{gordon:salmond:smith:1993} arises when
the joint proposal is chosen as,%
\begin{align*}
g(y_{t+1}|x_{t})\times g(x_{t+1}|x_{t})=1\times p(x_{t+1}|x_{t}),
\end{align*}%
in which case, $g(y_{t+1}|x_{t})$ is constant and $%
g(x_{t+1}|x_{t};y_{t+1})=p(x_{t+1}|x_{t})$. In this case, step (1) above
leaves the weights unchanged (as\ $\pi _{t|t+1}^{k}=\pi _{t}^{k}$).

The goal of the \apf is to get as close to full
adaption as possible,  when full adaption is not analytically possible. This is achieved
by making $g(y_{t+1}|x_{t})$ as close to $p(y_{t+1}|x_{t})$  as a function
of $x_{t}$ as possible (up to a constant of proportionality) and the density
$g(x_{t+1}|x_{t};y_{t+1})$ as close to $p(x_{t+1}|x_{t};y_{t+1})$ as possible.
Various procedures are found for doing this; see for example,
\cite{pitt:shephard:2001} and \cite{smith:santos:2006}.

The general ASIR estimator of $p(y_t|y_{1:t-1})$,  introduced and used by
\if0\blind{ \cite{pitt:2002}}, \fi{} \if1\blind { \cite{XXX}}\fi{}
is%
\begin{align}\label{asir_predlik_est}
\widehat{p}^{A}(y_{t}|y_{1:t-1})& =\left\{ \sum\limits_{k=1}^{M}\frac{\omega
_{t}^{k}}{M}\right\} \left\{ \sum\limits_{k=1}^{M}\omega
_{t-1|t}^{k}\right\} .
\end{align}%
The two sets of weights $\omega _{t}^{k}$ and $\omega _{t-1|t}^{k}$ are
defined above and calculated as part of the ASIR algorithm. These two
quantities are again a simple by-product of the algorithm. We define
the information in the swarm of particles at time $t$ as ${\cal A}%
_{t}=\{x_{t}^{k};\pi _{t}^{k}\}.$  For full adaption
$\omega _{t}^{k}=1$ and $\omega _{t-1|t}^{k}=p(y_{t}|x_{t-1}^{k})/M$ and
the first  summation in \eqref{asir_predlik_est} disappears.
For the SIR\ method, $\omega _{t}^{k}=p(y_{t}|x_{t}^{k})$ and $\omega
_{t-1|t}^{k}=\pi _{t-1}^{k}$ and  the second summation in (\ref%
{asir_predlik_est}) disappears.

{\normalsize The ASIR Algorithm \ref{alg:asir} is a flexible particle filter
approach when combined with stratification.  Theorem~\ref{thm: unbiased asir}
establishes that this algorithm together with the estimator of (\ref%
{asir_predlik_est}) is unbiased. This is important as it enables very
efficient likelihood estimators from the ASIR method to be used within an
MCMC algorithm.  }
\begin{thm}\label{thm: unbiased asir}
The ASIR likelihood
\begin{align} \label{eq: asir likelihood}
\widehat{p}^{A}(y_{1:t})& = \widehat{p}^{A}(y_{1})\prod_{t=2}^{T}\widehat{p}^{A}(y_{t}|y_{1:t-1})
\end{align}is unbiased in the sense that
\begin{align*}
E( \widehat{p}^{A}(y_{1:t}))& =p(y_{1:t})
\end{align*}
\end{thm}
The theorem is  proved in Section 7.4.2, Proposition 7.4.1 of \cite{delmoral:2004}. We give a more direct and accessible proof
in Appendix~\ref{A: unbiased likelihood}.

Our examples use the \spf and the \fapf, and the \papf  described in Appendix~\ref{app: partially adapted pf}.

\subparagraph{Simulated Likelihood}
The ASIR likelihood estimate~\ref{eq: asir likelihood} is called the simulated likelihood and Theorem~\ref{thm: unbiased asir} shows that the general ASIR particle filter provides a simulated likelihood that is an  unbiased estimate of the true likelihood function.
\citet{andrieu:doucet:holenstein:2010} show that we can view the simulated likelihood $\phat(y|\theta)$ as the density of $y$ conditional on $\theta $ and a set of auxiliary variables $u$ that are not a function of $\theta$ and such that
$\phat(y|\theta) = p_S(y|\theta, u ) $, where the subscript $S$ denotes a simulated likelihood,  and
\begin{align} \label{eq: unbiassed lik}
\int p_S(y|\theta,u)p(u)du  & = p(y|\theta).
\end{align}
The variables $u$ represent the uniform variates used for the multinomial/statified draws and the random variates (e.g. standard Gaussian) used in simulating from $g(x_{t+1}|x_t,y_{t+1})$.
It follows that the posterior $p_S(\theta|y)  = p(\theta|y)$ so that a method that simulates from $p_S(\theta,u|y)$ yields iterates from the correct posterior $p(\theta|y)$. We note that the ideas of using a simulated likelihood for Bayesian inference have also been explored, outside the area of particle
filters in the work of \cite{beaumont:2003:estimation} and
\cite{andrieu:2009:pseudo}
\bigskip

We note that the variance  of the log of the simulated likelihood is $O(T/M)$ so it will be necessary to take the number of particles $M = O(T)$ to keep a constant standard deviation as $T$ increases. This implies  that the \pf MCMC algorithm is of order $O(T^2)$ for $T$ large
and means that it is important to make the \pf as efficient as possible.



\section{Adaptive sampling for the simulated likelihood}\label{section:adaptive:sampling}
The target density for posterior inference is
$p_S(\theta,u|y) \propto p_S(y|\theta, u) p(\theta)p(u)$, where $p(\theta)$ is the prior for $\theta$. It may therefore be  possible to use a Metropolis-Hastings simulation method to generate samples from the target density as follows.
Suppose that given some initial $\theta_0$, the $j-1$ iterates $(\theta_1,u_1), \dots, (\theta_{j-1},u_{j-1})$ have been generated.
The $j$th iterate, $(\theta_{j},u_j)$, is generated from the proposal density $q_{j}(\theta;\thetat)p(u)$, which may also depend on some other value of $\theta$ which we call $\thetat$. Let $(\theta_{j}^p,u_j^p)$ be the proposed value of $(\theta_{j},u_j)$ generated from $q_{j}(\theta;\theta_{j-1})p(u).$
Then we take $(\theta_{j},u_j) = (\theta_{j}^p,u^p_j)$ with probability
\begin{align} \label{e:adaptive accep prob pf}
\alpha(\theta_{j-1},u_{j-1};\theta_{j}^p,u^{p}) = \min \biggl \{1,\frac{p_S(y|\theta_{j}^p,u_j^p)p(\theta_j^p)}
{p_S(y|\theta_{j-1},u_{j-1})p(\theta_{j-1})}
\dfrac{q_{j}(\theta_{j-1};\theta_j^p)}{q_{j}(\theta_j^p;\theta_{j-1})}\biggr \} \ ,
\end{align}
with $p(u_j^p) $ and $p(u_{j-1})$ cancelling out,
and take $(\theta_j,u_j) = (\theta_{j-1},u_{j-1})$ otherwise. We say that the proposal is independent if $q_{j}(\theta;\thetat)=q_{j}(\theta)$.

In adaptive sampling the parameters of $q_j(\theta;\thetat)$ are estimated from the iterates $\theta_1, \ldots, \theta_{j-2}$. When the likelihood can be measured exactly, i.e., in the non-particle filter case, then it can be shown that
under appropriate regularity conditions, the sequence of iterates $\theta_j, j \geq 1$,
converges to draws from the target distribution. See \cite{roberts:rosenthal:2007},
\cite{roberts:rosenthal:2008} and \citet{giordani:kohn:2010}.

Our article uses the \aimh scheme of \cite{giordani:kohn:2010}
and the \arwm scheme of \cite{roberts:rosenthal:2008}.
They are discussed in Appendix~\ref{app: adaptive sampling}.  Appendix~\ref{ss: convergence}
proves that the \aimh sampler converges to the target distribution under the given conditions.
Both the \spf and the \fapf satisfy these conditions almost automatically.
The appendix also shows that in the partially adapted case a
simple mixture of a \papf and the \spf also satisfies the conditions for convergence.


\subsection{Adaptive sampling and parallel computation}\label{ss: AS and parallel computation}
Carrying out Bayesian inference using the \pf{} and \MCMC simulation is computationally expensive. However, parallel processing can greatly increase the speed and therefore the
areas of application of our methods. For \aimh{} proposals we can use the following three step approach. Let $\theta^c$ the current value of $\theta$ generated by the sampling scheme and $q_c(\theta)$ the current proposal density for $\theta$. (a)~For each of $J$ processors generate  $K$ proposed values of $\theta$, which we write
as $\theta^{(p)}_{j,k},k=1, \dots, K$, and compute the corresponding logs
of the  ratios $ \phat (y|\theta^{(p)}_{j,k})p(\theta^{(p)}_{j,k}) /q(\theta^{(p)}_{j,k})$.
(b)~After each $K$ block of proposed values is generated for each processor,
carry out \MH{} selection of  the $JK$ proposed $\{\theta^{(p)}_{j,k}\}$
parameters using a single processor to  obtain $\{\theta_{j,k}\}$ draws from the chain.
This last step is fast because it is only necessary to draw uniform variates.
(c)~Use the previous iterates and the $\theta_{j,k}$ to update the proposal density $q_c(\theta)$ and $\theta_c$. In our applications of this approach, $K$ is chosen so that $KJ$ is aproximately the time between updates of the \aimh sampling scheme.

A second approach applies to all \MH sampling schemes, and in particular to the \arwm proposal. Suppose that $J$ processors are available.
The likelihood is estimated for a given $\theta$ on each of the processors using the particle filter with $M$ particles and these estimates are then averaged to get an estimate of the likelihood based on $JM$ particles. This approach is similar to, but faster, than using a single processor and makes it possible to estimate the likelihood using a large number of particles. However, for a given number of generated particles, the first  approach can be shown to be statistically more efficient than the second.

\subsection{Estimating the marginal likelihood}\label{subsection:bridge:sampling}
Marginal likelihoods are often used to compare two or more models.
For a given model, let $\theta$ be the vector of model parameters,
$p(y|\theta)$ the likelihood of the observations $y$ and $p(\theta)$ the prior for $\theta$.
The marginal likelihood is defined as
\begin{align}\label{eq:marg:likelihood}
p(y) & =\int p(y|\theta)p(\theta)d\theta.
\end{align}
which in our case can also be written as
\begin{align}\label{eq:marg:likelihood pf}
p(y) & =\int p_S(y|\theta,u)p(\theta)p(u) d\theta\ du \ .
\end{align}
It is often difficult to evaluate or estimate $p(y)$ in non-Gaussian state space models, although
auxiliary variable methods can be used in some problems. See \cite{fruwith:schnatter:wagner:2008}.
Appendix~\ref{A: BS and IS} briefly outlines how the marginal likelihood  can be estimated
using bridge or  importance sampling, with the computation carried out
within the adaptive sampling framework so that a separate simulation run is unnecessary.

\section{Comparing the standard SIR \pf with adapted ASIR \pf{s}} \label{S: comparing SIR with ASIR}
It is instructive to compare the performance of the \spf with the
fully and partially adapted \pf{s} for different signal to noise ratios and
using different numbers of particles.
We use two simulated examples. The first example compares the \spf and the \fapf
for a Gaussian  autoregressive signal observed with Gaussian noise.
This example is also of interest because we can compute the   exact likelihood
using the \KF, which is equivalent to using an infinite number of particles.
The second example compares the standard SIR to a \papf
using a a binomial model where we vary the signal to noise ratio by varying the number of binomial trials.

In both  examples, we make the  comparison in terms of three  criteria.
The first is the acceptance rate of the \aimh sampler, which we define
as the percentage of accepted draws.
The second  is the inefficiencies of the iterates of the parameters obtained using the \aimh{} method of \cite{giordani:kohn:2010}. The third is the standard deviation of the simulated log-likelihood $p_S(y|\theta, u)$ evaluated at the true value of $\theta$,
which is a good measure of how close the \pf likelihood is to the true likelihood $p(y|\theta)$.

We define the inefficiency of the sampling scheme for a given
parameter as the variance of the parameter estimate divided by its variance if the sampling scheme generates independent iterates. We estimate the inefficiency factor for a given parameter as $\text{IF} = 1+2\sum_{j=1}^{L^\star}\hat{\rho}_j,$
where $\hat{\rho}_j$ is the estimated autocorrelation of the parameter iterates
at lag $j$. As a rule of  thumb, the maximum number of lags $L^\star$ that
we use is $L^\star = \min\{ 1000, L\}$ , where $L$ is  the lowest index $j$ such that $|\hat{\rho}_j|<2/\sqrt{K}$ where $K$  is
the sample size used to compute $\hat{\rho}_j$. }

\subsection{Example~1: Autogressive model observed with noise} \label{SS: AR + noise}
Consider the following first order autoregression (AR(1)) plus noise model,
\begin{align}
y_{t} | x_{t}     &\sim \mathcal{N}(x_{t},\sigma^2) \nonumber \\
x_{t+1} | x_{t}   &\sim \mathcal{N}(\mu + \phi(x_{t}-\mu),\tau^2 ) \label{eq: ar1 state transition eqn}\\
x_0 &\sim \mathcal{N}(\mu,\tau^2/(1-\phi^2)). \label{eq: ar1 initial conditions}
\end{align}
The prior distributions are $\mu \sim \mathcal{N}(0,100), \phi \sim\mathcal{U}(0,1),
\sigma^2 \sim \mathcal{IG}(0.1,0.1)$ and
$\tau^2 \sim \mathcal{IG}(0.1,0.1)$. The notation $ \mathcal{N}(a,b^2) $ means a normal distribution with mean $a$ and variance $b^2$, $\mathcal{U}(a,b)$ means a uniform distribution on $(a,b)$ and $\mathcal{IG}(a,b)$ means  an inverse gamma distribution with shape parameter $a$ and scale $b$.

Our simulation study uses 50 replicated  data sets with 500 observations each, generated by setting $\mu=0$, $\phi=0.6$, $\tau^2=1$, $x_0\sim\mathcal{N}(\mu,\tau^2/(1-\phi^2))$ and two values for $\sigma^2 = \{0.01,1.0\}$, corresponding to high and low signal to noise ratios. We ran 30 000 iterations of the adaptive independent \MH{} for the posterior distribution using the standard particle filter and the fully adapted particle filter with differing number of particles. For completeness, we also ran the adaptive sampling scheme with the Kalman filter using the exact likelihood. The update times for the \aimh were at iterations  100, 200, 500, 1000, 1500, 2000, 3000, 4000, 5000, 10000, 15000 and 20000. We initialized the AIMH based on a normal proposal formed from 5000 draws of a previous run of the ARWM and the Kalman filter. For each signal to noise ratio we report the median MCMC parameter inefficiencies over the 50 replications as well as the interquartile range of the inefficiencies for differing numbers of particles. We also report results on
 the standard deviation of the simulated log likelihood for the standard particle filter and the \fapf. Specifically, for each of the 50 replicated data sets we computed the log likelihood at the true parameter values 1000 times for each of the two \pf{s} and obtained the median and the interquartile range of the medians and standard deviations of the log likelihoods across the 50 data replicates.


\subparagraph{Results for the high signal to noise case}\label{sss: ar1 high signal to noise}

Tables~\ref{table:ar1plusnoise:loglike:study:sigma2:001} and \ref{table:ar1plusnoise:MC:study:sigma2:001}  report the results for  the
high signal to noise case with $\sigma^2=0.01$.
Table~\ref{table:ar1plusnoise:loglike:study:sigma2:001} shows that the median variance of the simulated log likelihood $\log p_S(y|\theta,u)$ at the true parameter values for the \spf with 2000 particles is over 400 times higher than the median variance of the fully adapted \pf using 100 particles. This suggests that to get the same standard deviation for the simulated likelihood
we would need approximately 8000  times as many particles for the \spf as for the
\fapf as we know that the variance decreases approximately in inverse proportion to the number of particles.

Table~\ref{table:ar1plusnoise:MC:study:sigma2:001}
shows that the median acceptance rate of the \aimh sampler  using the \fapf
with 100 particles is about 1.75  times higher than the median acceptance rate of the standard \pf using 4000 particles. The table also shows that the median
parameter inefficiencies are about 3  times higher for the standard \pf using 4000 particles  than for the fully adapted \pf using  100 particles. Finally, the table shows that the \fapf using 500 particles performs almost as well as using the exact likelihood.

\begin{table}[!ht]
\centering
\caption{AR(1) + noise. High signal to noise.  Medians and interquantile ranges (IQR) of the estimated medians and standard deviations of the log of the simulated likelihood function at the true value for 50 different data sets.}
{\footnotesize
\begin{tabular}{l | cc | cc}\hline\hline
                  & \multicolumn{2}{c|}{Median}
                  & \multicolumn{2}{c}{Standard Deviation}\\\cline{2-5}
\up{N. Particles} & Median & IQR & Median & IQR \\\hline
        & \multicolumn{4}{c}{Standard Particle Filter}\\\hline
100     & -839.12 & 83.34 & 44.0381 & 21.0316 \\
500     & -729.43 & 26.09 & 10.5420 &  8.2875 \\
1000    & -719.10 & 20.13 &  5.5507 &  5.2818 \\
2000    & -714.95 & 18.16 &  2.8977 &  2.4716 \\
\hline  & \multicolumn{4}{c}{Fully Adapted Particle Filter}\\\hline
100     & -711.69 & 17.74 &  0.1431 &  0.0160 \\
\hline\hline
\end{tabular}
}
\label{table:ar1plusnoise:loglike:study:sigma2:001}
\end{table}

\begin{table}[!ht]
\centering
\caption{AR(1) + noise. High signal to noise. Medians and interquartile range (IQR) of the acceptance rates and the inefficiencies
over 50 replications of the autoregressive model using different particle filters and adaptive independent \MH{}.}
{\footnotesize
\begin{tabular}{l | rr | rr | rr | rr | rr }\hline\hline
& \multicolumn{2}{c|}{Ac. Rate}  & \multicolumn{2}{c|}{$\sigma^2$}
&\multicolumn{2}{c|}{$\tau^2$}&\multicolumn{2}{c|}{$\mu$}
&\multicolumn{2}{c}{$\phi$} \\\cline{2-11}
\up{N. Particles}  & Median & IQR & Median & IQR & Median & IQR & Median & IQR & Median & IQR \\\hline
&  \multicolumn{8}{c}{Kalman Filter}\\\hline
      & 72.18 &  4.89 &  1.93 &  0.34 &  1.85 &  0.35 &  1.76 &  0.20 &  1.83 &  0.24 \\
\hline&  \multicolumn{8}{c}{Standard Particle Filter}\\\hline
500   &   0.05 &  0.78 & 836.65 & 1727.31 & 1102.87 & 1732.95 & 1058.80 & 1733.35 & 979.64 & 1729.08 \\
1000  &   9.04 &  5.75 &  70.25 &   35.10 &   63.76 &   30.98 &   59.76 &   58.41 &  64.64 &   42.55 \\
2000  &  21.81 & 10.87 &  21.84 &   20.64 &   22.97 &   17.62 &   20.48 &   23.16 &  25.09 &   24.07 \\
4000  &  33.27 &  9.19 &   9.33 &    6.33 &    9.11 &    7.20 &    9.66 &    6.58 &   8.95 &    8.13 \\
\hline& \multicolumn{8}{c}{Fully Adapted Particle Filter}\\\hline
100   &  58.83 &  3.04 &   3.02 &    0.56 &    2.91 &    0.66 &    2.63 &    0.43 &   2.80 &    0.45 \\
500   &  67.64 &  2.30 &   2.23 &    0.31 &    2.08 &    0.31 &    2.02 &    0.20 &   2.10 &    0.24 \\
\hline\hline
\end{tabular}
}
\label{table:ar1plusnoise:MC:study:sigma2:001}
\end{table}

\subparagraph{Results for the low signal to noise case}\label{sss: ar1 low signal to noise case}
Tables~\ref{table:ar1plusnoise:loglike:study:sigma2:1} and \ref{table:ar1plusnoise:MC:study:sigma2:1}
report the results for  the low  signal to noise case with $\sigma^2=1.0$.
Table~ \ref{table:ar1plusnoise:loglike:study:sigma2:1} shows that the median variance of the log of the simulated likelihood at the true parameter values for the  \spf  using 1000 particles is about the same as the median variance of the simulated likelihood for the \fapf  using  100 particles, i.e. the variance of the simulated log likelihood of the \spf is about 10 times that of the \fapf for the same number of particles.

Table~\ref{table:ar1plusnoise:MC:study:sigma2:1} shows that the median
acceptance rates and parameter inefficiencies  of the \aimh{} sampler  using the fully adapted particle filter with 100 particles are about the those of the standard \pf using 4000 particles.

\begin{table}[!ht]
\centering
\caption{AR(1) + noise. Low  signal to noise. Medians and interquantile ranges (IQR) of the estimated medians and standard deviations of the log of the simulated likelihood function at the true value for 50 different data sets.}
{\footnotesize
\begin{tabular}{l | cc | cc}\hline\hline
                  & \multicolumn{2}{c|}{Median}
                  & \multicolumn{2}{c}{Standard Deviation}\\\cline{2-5}
\up{N. Particles} & Median & IQR & Median & IQR \\\hline
        & \multicolumn{4}{c}{Standard Particle Filter}\\\hline
100     & -904.0827 &  19.0013 &  2.4479 &  0.2372 \\
500     & -901.7877 &  18.8020 &  1.0793 &  0.1170 \\
1000    & -901.4966 &  18.7909 &  0.7629 &  0.0550 \\
\hline  & \multicolumn{4}{c}{Fully Adapted Particle Filter}\\\hline
100     & -901.4727 &  18.8540 &  0.7057 &  0.0398 \\
\hline\hline
\end{tabular}
}
\label{table:ar1plusnoise:loglike:study:sigma2:1}
\end{table}

\begin{table}[!ht]
\centering
\caption{AR(1) + noise. Low  signal to noise. Medians and interquartile range (IQR) of the acceptance rates and the inefficiencies
over 50 replications of the autoregressive model using different particle filters and adaptive independent \MH{}.}
{\footnotesize
\begin{tabular}{l | rr | rr | rr | rr | rr }\hline\hline
& \multicolumn{2}{c|}{Ac. Rate}  & \multicolumn{2}{c|}{$\sigma^2$}
&\multicolumn{2}{c|}{$\tau^2$}&\multicolumn{2}{c|}{$\mu$}
&\multicolumn{2}{c}{$\phi$} \\\cline{2-11}
\up{N. Particles}  & Median & IQR & Median & IQR & Median & IQR & Median & IQR & Median & IQR \\\hline
&  \multicolumn{8}{c}{Kalman Filter}\\\hline
      & 73.81 &  2.26 &  2.04 &  0.38 &  2.12 &  0.46 &  1.73 &  0.12 &  1.94 &  0.25 \\
\hline&  \multicolumn{8}{c}{Standard Particle Filter}\\\hline
500   &  6.84 &  6.28 & 92.85 & 76.67 & 95.95 & 71.78 & 71.63 & 51.26 & 85.93 & 68.60 \\
1000  & 29.86 & 18.15 & 19.67 & 27.30 & 21.10 & 24.91 & 10.87 & 16.07 & 18.18 & 28.90 \\
2000  & 42.47 & 13.37 & 11.87 & 11.20 & 10.72 &  8.98 &  5.36 &  3.62 &  9.02 &  6.88 \\
4000  & 52.95 & 11.85 &  6.38 &  6.86 &  6.16 &  4.53 &  3.28 &  2.54 &  5.02 &  4.11 \\
\hline& \multicolumn{8}{c}{Fully Adapted Particle Filter}\\\hline
100   & 53.94 &  9.24 &  3.48 &  0.88 &  3.53 &  0.84 &  3.27 &  1.32 &  3.62 &  0.92 \\
\hline\hline
\end{tabular}
}
\label{table:ar1plusnoise:MC:study:sigma2:1}
\end{table}

\subsection{Example 2: Binomial model with an autoregressive state equation}
\label{ss: binomial model}
Consider observations generated from the following dynamic binomial model
\begin{align*}
y_t & \sim \Bin(m,\pi_t) \ , \quad
\pi_t  = \exp(x_t)/(1+\exp(x_t))
\end{align*}
where $m$ is the number of trials and $\pi_t$ is the probability of success of each trial.
The states $x_t$ follow the first order AR(1) model \eqref{eq: ar1 state transition eqn} whose initial distribution is \eqref{eq: ar1 initial conditions}.
The prior distributions for the parameters are $\mu \sim \mathcal{N}(0,100),
\phi \sim \mathcal{U}(0,1)$ and $\tau^2 \sim \mathcal{HN}(100)$. We use the notation $\mathcal{HN}(b^2)$ to mean a half-normal distribution with scale $b$.

Our simulation study is organized similarly to that in Section~\ref{SS: AR + noise}.
The data generating process takes
$\mu=0$, $\phi=0.97$, $\tau^2=0.25$ and the number of trials takes the two values $m=\{100,500\}$. We initialized the AIMH based on a normal proposal formed from 5000 draws of a previous run of the ARWM and the \papf using 500 particles.

We note the following about the binomial density.
\begin{enumerate}
\item
The \spf will do worse as the number of trials $m$ increases
because  the the measurement density becomes
more informative and peaked so  the variance of the  weights increases.

\item The opposite is true for the \papf  method. The measurement density
$p(y_t|x_t)$ tends to normality as $m$ increases by the \clt so
that the \papf tends to a \fapf.
 Hence the \papf method actually improves as $m$
becomes larger and this is seen in
Tables~\ref{table:binomial:loglike:study:m500} and
\ref{table:binomial:loglike:study:m100} below.
\end{enumerate}

\subparagraph{High signal to noise case}\label{sss: binom high signal to noise}
Tables~\ref{table:binomial:loglike:study:m500} and \ref{table:binomial:data:MC:study:m500}
report the results for the high signal to noise case, with the number of trials set at $m = 500$.
Table~\ref{table:binomial:loglike:study:m500}  shows that the variance of the simulated log likelihood  at the true parameter values for the standard  \pf with 4000 particles is  about 2.5  times higher than that of the \papf  when 100 particles are used. This means that it is necessary to have 100 times as many particles using the \spf to get the same noise level for the simulated log likelihood as for the \papf.

Table~\ref{table:binomial:data:MC:study:m500} shows that the median
acceptance rate of the \aimh{} sampler  using the \papf with 100 particles is about 1.3 times higher than the median acceptance rate of standard \pf  using 4000 particles.
The table also shows that the median
parameter inefficiencies are 1.5 times higher for the standard \pf using 4000 particles
than they are for the \papf using 100 particles.

\begin{table}[!ht]
\centering
\caption{Binomial example. High signal to noise.  Medians and interquantile ranges (IQR) of the estimated medians and standard deviations of the log-likelihood function at the true value for 50 different data sets.}
{\footnotesize
\begin{tabular}{l | cc | cc}\hline\hline
                  & \multicolumn{2}{c|}{Median}
                  & \multicolumn{2}{c}{Standard Deviation}\\\cline{2-5}
\up{N. Particles} & Median & IQR & Median & IQR \\\hline
        & \multicolumn{4}{c}{Standard Particle Filter}\\\hline
500     & -2441.29 & 114.63 &  3.3149 &  1.5044 \\
1000    & -2439.64 & 114.39 &  2.0737 &  0.8711 \\
2000    & -2438.88 & 117.55 &  1.4106 &  0.3203 \\
4000    & -2438.45 & 118.61 &  0.9478 &  0.1593 \\
\hline  & \multicolumn{4}{c}{Partially Adapted Particle Filter}\\\hline
100     & -2438.28 & 119.38 &  0.6182 &  0.1358 \\
\hline\hline
\end{tabular}
}
\label{table:binomial:loglike:study:m500}
\end{table}

\begin{table}[!ht]
\centering
\caption{Binomial example. High signal to noise. Medians and interquartile range (IQR) of the acceptance rates and the inefficiencies
over 50 replications of the binomial model using different particle filters and adaptive independent \MH.}
{\footnotesize
\begin{tabular}{l | rr | rr | rr | rr }\hline\hline
& \multicolumn{2}{c|}{Ac. Rate}  & \multicolumn{2}{c|}{$\mu$}
&\multicolumn{2}{c|}{logit($\phi$)}&\multicolumn{2}{c}{$\log(\tau^2)$} \\\cline{2-9}
\up{N. Particles}  & Median & IQR & Median & IQR & Median & IQR & Median & IQR \\\hline
&  \multicolumn{8}{c}{Standard Particle Filter}\\\hline
500   &   3.66 &  2.57 & 92.85 & 99.02 &100.32 &369.21 &105.76 &330.84 \\
1000  &  13.28 &  4.13 & 39.84 & 47.49 & 38.77 & 32.71 & 40.89 & 23.31 \\
2000  &  27.18 &  5.54 & 12.26 &  7.83 & 12.15 &  7.61 & 13.29 & 10.00 \\
4000  &  39.73 &  5.85 &  6.01 &  3.86 &  5.93 &  3.00 &  5.82 &  2.14 \\
\hline& \multicolumn{8}{c}{Partially Adapted Particle Filter}\\\hline
100   &  51.52 & 11.17 &  3.82 &  3.06 &  3.74 &  2.09 &  3.39 &  1.35 \\
\hline\hline
\end{tabular}
}
\label{table:binomial:data:MC:study:m500}
\end{table}

\subparagraph{Low signal to noise case}
Tables~\ref{table:binomial:loglike:study:m100} and \ref{table:binomial:data:MC:study:m100}
report the results for the low signal to noise case, with the number of trials set at $m = 100$.
Table~\ref{table:binomial:loglike:study:m100}  shows that the median variance of simulated log likelihood  at the true parameter values for the   \spf with 4000 particles is  about the same as that obtained by the \papf using 250 particles, i.e. the \spf requires about 16 times as many particles to obtain the same standard deviation as the \papf.

Table~\ref{table:binomial:data:MC:study:m100} shows that the median
acceptance rate of the \aimh{} sampler  using the \papf with 200 particles is
higher than the median acceptance rate of standard \pf  using 2000 particles.
The table also shows that the median
parameter inefficiencies for the standard \pf using 2000 particles are higher than the median inefficiencies for  the \papf using 200 particles.

\begin{table}[!ht]
\centering
\caption{Binomial example. Low signal to noise.  Medians and interquantile ranges (IQR) of the estimated medians and standard deviations of the log-likelihood function at the true value for 50 different data sets.}
{\footnotesize
\begin{tabular}{l | cc | cc}\hline\hline
                  & \multicolumn{2}{c|}{Median}
                  & \multicolumn{2}{c}{Standard Deviation}\\\cline{2-5}
\up{N. Particles} & Median & IQR & Median & IQR \\\hline
        & \multicolumn{4}{c}{Standard Particle Filter}\\\hline
500     & -1729.05 & 109.49 & 1.8385 & 0.2960 \\
1000    & -1728.16 & 109.63 & 1.2704 & 0.2457 \\
2000    & -1727.69 & 109.72 & 0.8827 & 0.1458 \\
4000    & -1727.50 & 109.82 & 0.6300 & 0.0711 \\
\hline  & \multicolumn{4}{c}{Partially Adapted Particle Filter}\\\hline
100     & -1727.81 & 109.84 & 0.9867 & 0.1074 \\
200     & -1727.55 & 109.85 & 0.7132 & 0.0810 \\
500     & -1727.41 & 109.90 & 0.4465 & 0.0550 \\
\hline\hline
\end{tabular}
}
\label{table:binomial:loglike:study:m100}
\end{table}

\begin{table}[!ht]
\centering
\caption{Binomial example. Low  signal to noise. Medians and interquartile range (IQR) of the acceptance rates and the inefficiencies
over 50 replications of the binomial model using different particle filters and adaptive independent \MH{}.}
{\footnotesize
\begin{tabular}{l | rr | rr | rr | rr }\hline\hline
& \multicolumn{2}{c|}{Ac. Rate}  & \multicolumn{2}{c|}{$\mu$}
&\multicolumn{2}{c|}{logit($\phi$)}&\multicolumn{2}{c}{$\log(\tau^2)$} \\\cline{2-9}
\up{N. Particles}  & Median & IQR & Median & IQR & Median & IQR & Median & IQR \\\hline
&  \multicolumn{8}{c}{Standard Particle Filter}\\\hline
500   &  17.02 &  4.82 & 29.37 & 17.65 & 31.85 & 16.33 & 31.57 & 15.90 \\
1000  &  31.01 &  5.46 & 10.85 &  7.54 & 10.51 &  7.14 & 10.33 &  4.73 \\
2000  &  43.47 &  5.91 &  5.27 &  3.30 &  5.17 &  2.10 &  5.00 &  1.79 \\
4000  &  54.02 &  4.91 &  3.70 &  3.35 &  3.48 &  1.94 &  3.03 &  0.58 \\
\hline& \multicolumn{8}{c}{Partially Adapted Particle Filter}\\\hline
100   &  39.16 &  7.15 &  5.06 &  3.60 &  5.99 &  2.63 &  5.96 &  2.50 \\
200   &  50.59 &  5.19 &  3.62 &  2.26 &  3.81 &  2.02 &  3.59 &  1.21 \\
500   &  60.80 &  4.19 &  2.80 &  3.32 &  2.47 &  1.23 &  2.32 &  0.32 \\
\hline\hline
\end{tabular}
}
\label{table:binomial:data:MC:study:m100}
\end{table}

\section{Performance of the adaptive sampling schemes on real examples}\label{section:algorithm:comparison}
This section uses real data to illustrate the flexibility
and wide applicability of the approach that combines particle filtering with adaptive sampling. All that is necessary for model estimation and model comparison by marginal likelihood is to code up a \pf to evaluate the simulated likelihood and to code up the prior on the parameters. We also illustrate the difference in performance between the \arwm sampling scheme of \cite{roberts:rosenthal:2008} and that  the \aimh scheme of \cite{giordani:kohn:2010}.
This comparison is interesting for two reasons.
First, the \aimh scheme tries to obtain a good approximation to the posterior density, whereas the \arwm aims for some target acceptance rate. Second, we claim that
any \imh scheme (of which the \aimh scheme of \cite{giordani:kohn:2010} is an example),
is more suitable  to be implemented in parallel than a \MH scheme with a
proposal that depends on the previous iterate (of which the \arwm
scheme of \cite{roberts:rosenthal:2008} is an example).

The comparison between the two schemes is in terms of three criteria. The first two are
the acceptance rate of the Metropolis-Hastings method and the
inefficiency factors (IF) of the parameters and
are defined  in Section~\ref{S: comparing SIR with ASIR};
they  are independent of the way the algorithms are implemented.
However, these two criteria do not take into account the times taken by the samplers. To obtain an overall measure of the effectiveness of a sampler, we define its equivalent computing time $ ECT = 1000 \times IF \times t$, where $t$ is the time per iteration of the sampler. We interpret $ECT$ as the time taken by the sampler to attain the same accuracy as that attained by 1000 independent draws of the same sampler. For two samplers $a$ and $b$,  $ECT_a/ECT_b$ is the ratio of times taken by them to achieve the same accuracy.
We note that the time per iteration for a given sampling algorithm depends on how it is
implemented, i.e. the language used, whether operations are vectorized, etc.

The results  presented are for a single processor and the two parallel
methods discussed in Section~\ref{ss: AS and parallel computation}.
To simplify the presentation, we mainly present results for the \spf.

\subsection{Example~1: Stochastic volatility model with leverage and outliers} \label{ss: sv model}
The first example considers the univariate stochastic volatility (SV) model
\begin{align}\label{eq: sv general}
\begin{split}
y_{t}      & = K_t\exp(x_t/2)\varepsilon_t,\hspace{2cm}\varepsilon_t\sim\mathcal{N}(0,1) \\
x_{t+1} & =  \mu + \phi(x_{t}-\mu)+\sigma_\eta\eta_t,\hspace{0.6cm}\eta_t\sim\mathcal{N} (0,1)
\end{split}
\end{align}
where $\text{corr}(\varepsilon_t,\eta_t)=\rho$, $\Pr(K_t=2.5)=\omega$ and $\Pr(K_t=1)=1-\omega$, with $\omega << 1$. This is a state space model with a non-Gaussian observation equation  and a Gaussian state transition equation for the latent volatility $x_t$ which follows a first order autoregressive model. The SV model allows for leverage because the errors in the observation and state transition equations can be correlated. The model also allows for outliers in the observation equation because the standard deviation of $y_t$ given $x_t$ can be 2.5 its usual size when $K_t = 2.5$. To complete the model specification,  we assume that all parameters are independent a priori with the following prior distributions: $\mu\sim\mathcal{N}(0,10^2)$, $\phi\sim\mathcal{TN}_{(0,1)}(0.9,0.1)$, $\sigma_\eta^2\sim\mathcal{IG}(0.01,0.01)$, and $\rho\sim\mathcal{TN}_{(-1,1)}(0,10^6)$.
We use the notation  $\mathcal{TN}_{(c,d)}(a,b)$ to mean a truncated normal
with location $a$ and scale $b$ restricted to the interval $(c,d)$  and $\mathcal{IG}(a,b)$ is an inverse gamma distribution with shape parameter $a$, scale parameter $b$ and mode $b/(a+1)$. We set $\omega=0.03$ in the general model to indicate that outliers are rare apriori.

\cite{shephard:2005} reviews SV models and a model of the form \eqref{eq: sv general} is estimated by \if0\blind{ \citet{malik:pitt:2008}}\fi{} \if1\blind{\cite{WWW}}\fi
by maximum likelihood using the smooth \pf{}.

\subparagraph{S\&P 500 index}\label{sss: sp}
We apply the  SV model \eqref{eq: sv general} to the Standard and Poors (S\&P) 500 data from 02/Jan/1970 to 14/Dec/1973 obtained from Yahoo Finance web site\footnote{ http://au.finance.yahoo.com/q/hp?s=\^{}GSPC}. The data consists of $T=1~000$ observations.

Table~\ref{table:SV:SP500:MC:study:K500} shows the acceptance rates, the inefficiencies and the equivalent computing time over 10 replications of the stochastic volatility model using the\spf and the two  adaptive \MH{} schemes. The analysis uses the SV model without leverage or outliers.
In the table, SP stands for a single processor, MP$_1$ for multiprocessor method 1 and MP$_2$ for multiprocessor method 2 (where the simulated likelihood is obtained as an average) described in Section~\ref{ss: AS and parallel computation}.
We use eight processors for both the  MP$_1$ and MP$_2$ schemes.
The  basic number of particles in this example is $K=500$, which means
that  SP uses  4000 particles in a single processor,  MP$_1$ uses
4000 particles in each processor and MP$_2$
uses 500 particles in each processor. We ran all the algorithm for 10000 iterations and took the last 5000 to compute the results. The equivalent computing time is obtained by taking the overall time divided by the number of iterations times the inefficiency times 1000. The update times for the \aimh using SP or MP$_2$ were at 100, 200, 500, 1000, 2000, 3000, 4000, 5000, 6000 and 7500. The block sizes (also the update times) for the adaptive \MH{} MP$_1$ were 15, 25, 60, 125, 250, 375, 500, 625, 750 and 940.

The table shows that the acceptance rates of the \aimh sampler are about twice those of the \arwm and the inefficiencies are about 1/6 of those for the \arwm. For ECT, the best  approach for \aimh is MP$_1$ and  is between 11 and 18 times better than the best available  approach (which is MP$_2$) for \arwm. Qualitatively similar results were obtained for $K = 1000$ particles.

\begin{table}[!ht]
\centering
\caption{SV model. Medians and interquartile range (between brackets) of the acceptance rates, the inefficiencies and the equivalent computing time (t$\times$inefficiency$\times$1000) over 10 replications of the stochastic volatility model using the standard particle filter and differing adaptive \MH{} schemes. SP = single processor, MP$_1$ = multiprocessor \MH{} and MP$_2$ = multiprocessor averaging the likelihood function.}
{\footnotesize
\begin{tabular}{l | c | ccc | ccc }\hline\hline
&  & \multicolumn{3}{c|}{Inefficiency} & \multicolumn{3}{c}{Equivalent Computing Time}\\\cline{3-8}
\up{Algorithm}  &\up{Ac. Rate}&  $\mu$  & $\text{logit}(\phi)$ & $\log(\sigma_\eta^2)$
                              &  $\mu$  & $\text{logit}(\phi)$ & $\log(\sigma_\eta^2)$ \\\hline
                &  24.5  &  25.47  &  30.20  &  20.00  &  13463.7 & 15972.9  & 10603.0  \\
\up{ARWM-SP}    &  (0.6) & (16.33) &  (5.12) &  (2.80) & (8589.8) & (2655.3) & (1514.0) \\ \hline
                &  22.2  &  25.04  &  31.07  &  20.51  &  3594.5  &  4453.4  &  2930.4  \\
\up{ARWM-MP$_2$}&  (5.5) & (12.96) & (14.02) &  (8.83) & (1886.9) & (1994.3) & (1284.3) \\\hline
                &  51.6  &   6.45  &   3.46  &   3.08  &  3430.9  &  1841.7  &  1638.7  \\
\up{AIMH-SP}    &  (0.8) &  (7.12) &  (0.83) &  (0.07) & (3777.9) &  (439.7) &   (39.9) \\\hline
                &  52.3  &   3.44  &   3.42  &   3.63  &   237.8  &   235.3  &   250.0  \\
\up{AIMH-MP$_1$}&  (3.3) &  (2.13) &  (0.76) &  (0.50) &  (147.4) &   (50.5) &   (34.6) \\\hline
                &  53.6  &   4.40  &   3.52  &   3.15  &   627.1  &   504.4  &   451.1  \\
\up{AIMH-MP$_2$}&  (3.2) &  (6.74) &  (3.37) &  (0.56) &  (975.4) &  (482.5) &   (75.8) \\
\hline\hline
\end{tabular}
}
\label{table:SV:SP500:MC:study:K500}
\end{table}

\subparagraph{Model selection}
We now use  importance sampling and bridge sampling to compute the marginal likelihoods of  the four SV models: the model with no leverage effect ($\rho = 0$) and no outlier effect ($\omega = 0$), the model that allows for leverage but not outliers, the model that allows for outliers but no leverage and the general model that allows for both outliers and leverage.
Table \ref{table:SV:model:comparison} shows the logarithms of the marginal likelihoods of the four models for a single run of each algorithm. The differences between the two approaches are very small. In this example, and based on our prior distributions, the SV model with leverage effects but no outliers has the highest marginal likelihood.

\subparagraph{Posterior estimates of model parameters}
The estimated posterior means and standard deviations of all four models are given in Table~\ref{table:SP500:summary:stats}.

\begin{table}[!ht]
\centering
\caption{Logarithms of the marginal likelihoods for four different SV models for the two \pf{}
algorithms computed using the \aimh{} algorithm. $BS$ and $IS$ mean bridge sampling and importance sampling.}
{\footnotesize
\begin{tabular}{ l | cc | cc}\hline\hline
          & \multicolumn{2}{c|}{Standard Particle Filter}
       &    \multicolumn{2}{c}{Partially Adapted Particle Filter}\\\cline{2-5}
 \up{Model}   &	$\log(p_{BS}(y))$ & $\log(p_{IS}(y))$ &	$\log(p_{BS}(y))$ & $\log(p_{IS}(y))$ \\\hline
SV           & -1072.9    &   -1072.9 & -1072.9   &   -1072.9  \\
SV Lev.      & -1065.0    &   -1065.0 & -1065.0   &   -1065.0  \\
SV Out.      & -1076.6    &   -1076.6 & -1076.5   &   -1076.4  \\
SV Lev. Out. & -1069.3    &   -1069.3 & -1069.2   &   -1069.3  \\
\hline\hline
\end{tabular}
}
\label{table:SV:model:comparison}
\end{table}
\begin{table}[!ht]
\centering
\caption{S\&P 500 data: Estimated posterior means and standard deviations for all four stochastic volatility models.}
{\footnotesize
\begin{tabular}{l | rr | rr | rr | rr}\hline \hline
Parameter &\multicolumn{2}{c|}{SV}&\multicolumn{2}{c|}{SV Lev.}
&\multicolumn{2}{c|}{SV Out.}&\multicolumn{2}{c}{SV Lev. Out}\\\cline{2-9}
          &   Mean  & S. Dev. &   Mean  & S. Dev.&   Mean  &  S. Dev.&    Mean  & S. Dev.\\\hline
$\mu$     & -0.4329 &  1.2314 & -0.5642 & 0.1500 & -0.1786 &  2.2812 &  -0.5756 & 0.3497 \\
$\phi$    &  0.9879 &  0.0097 &  0.9811 & 0.0063 &  0.9907 &  0.0086 &   0.9830 & 0.0065 \\
$\tau^2$  &  0.0142 &  0.0068 &  0.0106 & 0.0037 &  0.0116 &  0.0053 &   0.0091 & 0.0034 \\
$\rho$    & --------& --------& -0.7608 & 0.0960 & --------& --------&  -0.7652 & 0.0960 \\
\hline \hline
\end{tabular}
}
\label{table:SP500:summary:stats}
\end{table}


\subsection{Example 2: GARCH model observed with noise}\label{SS: garch with noise}
The GARCH(1,1) model is used extensively to model financial returns, see for example,
\cite{bollerslev:engle:nelson:1994}.
In this section we consider the GARCH(1,1) model observed with Gaussian noise which is a more flexible version of the basic model. The model is,
\begin{align*}
y_{t} | x_{t}        &\sim \mathcal{N}(x_{t},\tau^2)\\
x_{t+1} | \sigma_{t+1}^2   &\sim \mathcal{N}(0,\sigma_{t+1}^2 ) \\
\sigma_{t+1}^2       &= \alpha + \beta x_t^2 + \gamma \sigma_t^2\\
x_0               &\sim  \mathcal{N}(0,\alpha/(1-\beta-\gamma)).
\end{align*}
The priors on $\tau^2$ and $\alpha$ are
$\tau^2 \sim\mathcal{HN}(100)$ and $ \alpha  \sim\mathcal{HN}(100)$.
The joint prior for $\beta$ and $\gamma$ is uniform in the region $\beta> 0, \gamma > 0, \beta + \gamma < 1$.

It is straightforward to show that this model is fully adapted. See Appendix~\ref{app: fully adaptive pf}
Instead of using the GARCH(1,1) model with noise we can use other members of the GARCH family, e.g. an EGARCH process observed with noise. All such models are fully adapted.

\begin{figure}[!ht]
  \centering
  \includegraphics[width=9cm,angle=0]{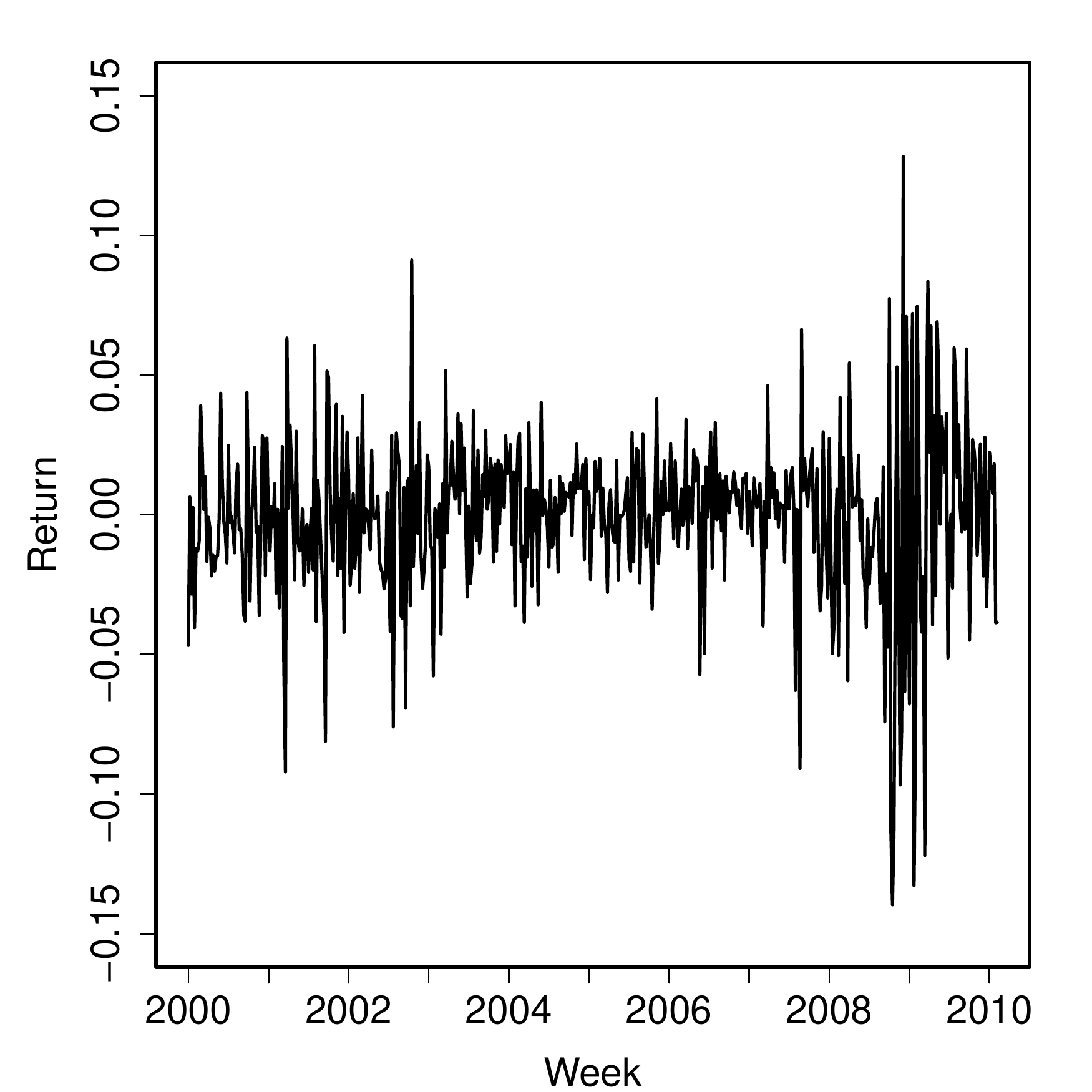}\\
  \caption{UK MSCI weekly returns from 6 January 2000 to 28 January 2010.}
  \label{fig:UK:returns}
\end{figure}

\subparagraph{MSCI UK index returns} We model the weekly MSCI UK index returns from 6 January 2000 to 28 January 2010 corresponding to 526 weekly observations
shown in Figure~\ref{fig:UK:returns}.

Table \ref{table:GARCH:UK:index} compares different adaptive sampling algorithms and particles filters in terms of acceptance rates, inefficiencies and equivalent computing times. Medians and interquantile ranges are computed using 50 replication of each adaptive sampling, particle filter and number of particles. The adaptive sampling algorithms were run for 30 000 iterations. The first 20 000 draws were discarded and the remainder used to compute the statistics. The update times for the adaptive independent \MH{} were at 100, 200, 500, 1000, 1500, 2000, 3000, 4000, 5000, 10000, 15000  and 20000. The initial value and proposal distribution for all algorithms were based on a single short run of the adaptive random walk and all results are for a single processor.

The table shows that  the \fapf is much more efficient than the \spf for both the \arwm and the \aimh samplers, and that the \aimh sampler is much more efficient than the \arwm sampler for both the standard and the \fapf{s}. In particular, the table shows that \aimh combined with full adaptation using 200 particles is about four times as efficient as the \arwm sampler using the \spf and 10, 000 particles.

\begin{table}[!ht]
\centering
\caption{Medians and interquartile range (between brackets) of the acceptance rates, the inefficiencies and
the equivalent computing time for 50 replications of the Gaussian GARCH model observed with noise applied to
the UK index return using differing particle filters, number of particles and adaptive \MH{} algorithms. A single processor was used for all results. }
{\footnotesize
\begin{tabular}{l | l | c | cccc }\hline\hline
               & \# of     & Accept. & \multicolumn{4}{c}{Inefficiency}  \\\cline{4-7}
\up{Algorithm} & Particles & Rate    & $\tau^2$ & $\alpha$ & $\beta$ & $\gamma$ \\
\hline&  \multicolumn{6}{c}{Standard Particle Filter}\\\hline
         &           &  6.90  &  82.75  &  93.79  &  89.87  &  90.84   \\
         &\up{1000}  & (0.99) & (15.40) & (19.02) & (18.98) & (24.55) \\\cline{2-7}
         &           & 14.72  &  39.36  &  48.69  &  56.63  &  53.00 \\
\up{ARWM}&\up{5000}  & (1.07) &  (5.96) & (15.40) & (11.01) & (14.48) \\\cline{2-7}
         &           & 17.11  &  37.93  &  43.27  &  50.71  &  47.87  \\
         &\up{10000} & (1.35) & (11.42) &  (6.85) &  (9.73) & (16.48) \\\hline
         &           & 13.25  &  40.47  &  46.70  &  44.14  &  42.32   \\
         &\up{1000}  & (1.94) & (24.91) & (26.91) & (14.51) & (20.83) \\\cline{2-7}
         &           & 29.97  &   8.81  &  10.87  &  11.68  &  11.82  \\
\up{AIMH}&\up{5000}  & (2.23) &  (3.49) &  (2.94) &  (2.94) &  (3.48)  \\\cline{2-7}
         &           & 33.64  &   7.42  &   9.24  &   9.08  &  10.07  \\
         &\up{10000} & (1.51) &  (1.46) &  (2.25) &  (3.26) &  (6.39) \\
\hline&  \multicolumn{6}{c}{Fully Adapted Particle Filter}\\\hline
         &           & 15.35  &  43.82  &  48.42  &  56.42  &  49.56   \\
         &\up{200}   & (1.04) &  (8.54) & (10.54) & (10.07) & (15.87)  \\\cline{2-7}
         &           & 17.45  &  36.52  &  43.21  &  52.31  &  48.06 \\
\up{ARWM}&\up{500}   & (1.23) &  (7.25) & (12.05) & (13.67) & (15.76) \\\cline{2-7}
         &           & 18.26  &  34.06  &  40.35  &  51.91  &  45.56   \\
         &\up{1000}  & (1.22) &  (5.88) &  (7.70) &  (7.38) & (17.74) \\\hline
         &           & 30.28  &  10.16  &  11.72  &  11.34  &  10.44  \\
         &\up{200}   & (1.61) &  (4.44) &  (4.39) &  (3.46) &  (3.33)  \\\cline{2-7}
         &           & 34.44  &   7.76  &   8.68  &   8.53  &  10.17 2\\
\up{AIMH}&\up{500}   & (1.38) &  (1.97) &  (3.27) &  (2.02) &  (5.40) \\\cline{2-7}
         &           & 36.19  &   6.61  &   9.51  &   7.90  &   8.01  \\
         &\up{1000}  & (1.54) &  (2.05) &  (3.68) &  (1.66) &  (5.14) \\
\hline\hline
\end{tabular}
}
\label{table:GARCH:UK:index}
\end{table}

Table~\ref{table:GARCH:UK:index:SD:LLK} shows the standard deviations of the simulated log-likelihood function for the particle filters using differing number of particles. The statistics are based on 1000 replications of the particle filter with the parameters fixed at their posterior means. The summary statistics of the posterior distribution is shown in Table \ref{table:summary:stats:GARCH:UK}. The table shows that the standard deviation of the simulated log likelihood using the \fapf with 500 particles is smaller than the standard deviation of the simulated log likelihood using 10, 000 particles.
\begin{table}[!ht]
\vspace{3mm}
\centering
\caption{UK MSCI index returns: Standard deviation of the simulated
log-likelihood function at the posterior mean for \spf and \fapf using various numbers of particles and 1000 replications.}
{\footnotesize
\begin{tabular}{l | r| l| r}\hline\hline
{\#  Particles}  & SIR sd & {\# Particles}  & FAPF sd  \\\hline
1000    &   1.73 & 200 & 0.67\\
5000    &   0.71 & 500 & 0.43\\
10000   &   0.51 & 1000 & 0.31 \\
\hline\hline
\end{tabular}
}
\label{table:GARCH:UK:index:SD:LLK}
\end{table}

Table~\ref{table:summary:stats:GARCH:UK} is a summary of the posterior distributions of the four parameters.

\begin{table}[!ht]
\centering
\caption{Summary of statistics of the posterior distribution.}
{\footnotesize
\begin{tabular}{l | cc }\hline\hline
Parameter &   Mean     &  St.Dev.   \\\hline
$\tau^2$&  0.0002700 &  0.0000462 \\
$\alpha$  &  0.0000495 &  0.0000289 \\
$\beta$   &  0.8927539 &  0.0672126 \\
$\gamma$  &  0.0377854 &  0.0412842 \\
\hline\hline
\end{tabular}
}
\label{table:summary:stats:GARCH:UK}
\end{table}

\if0\blind
{
\section*{Acknowledgment}
The research of Robert Kohn and Ralph S. Silva was partially supported by an ARC Discovery Grant DP0667069}
\fi

\bibliographystyle{asa}
\bibliography{pf}

\begin{appendices}
\section{Proof that the AISR likelihood is unbiased}\label{A: unbiased likelihood}
This appendix proves Theorem~\ref{thm: unbiased asir}
using an iterated expectations argument on the simulated likelihood.
A similar result is obtained
in Proposition 7.4.1 of Section 7.4.2 by \cite{delmoral:2004} by showing that the
difference of the measure on the states induced by the particle
filter and that of the limiting Feyman-Kac
measure is a martingale.
We believe that our proof which deals specifically with the unbiasedness of the simulated
likelihood is simpler and more direct and is accessible to a much wider range of
readers.


Before giving the proof we define some terms that are used
in Algorithm~\ref{alg:asir},
\begin{align*}
\widehat{p}_{M}^{A}(x_{t}|y_{1:t})= & \sum_{k=1}^{M}\pi _{t}^{k}\delta
(x_{t}-x_{t}^{k}), \text{        } \text{where }\pi _{t}^{k}\text{ is given in Step (4).}
\\
\widehat{g}_{M}^{A}(x_{t}|y_{1:t+1})= & \sum_{k=1}^{M}\pi _{t|t+1}^{k}\delta
(x_{t}-x_{t}^{k}),\text{ } \text{      }  \text{where }x_{t}^{k}\sim \widehat{p}%
_{M}^{A}(x_{t}|y_{1:t}), \\
\widehat{g}_{M}^{A}(x_{t}|y_{1:t})= & \int g(x_{t}|\widetilde{x}_{t-1};y_{t})%
\widehat{g}_{M}^{A}(\widetilde{x}_{t-1}|y_{1:t})d\widetilde{x}_{t-1}, \\
\omega_{t|t_+1}(x_t)  = & g(y_{t+1}|x_t) \pi_t \ , \text{      }  \omega_{t+1}(x_{t+1};x_t)  = \frac{p(y_{t+1}|x_{t+1}) p(x_{t+1}|x_t) }{g(y_{t+1}|x_t)g(x_{t+1}|x_t,y_{t_+1})}
\end{align*}
The term $\phat_{M}^{A}(x_{t}|y_{1:t})$ is the
empirical filtering density arising from step 4 of
Algorithm~\ref{alg:asir}. The second term
$\widehat{g}_{M}^{A}(x_{t}|y_{1:t+1})$, is the empirical \textquotedblleft
look ahead" approximation drawn from in step 2. The expression $\widehat{g}%
_{M}^{A}(x_{t}|y_{1:t})$ is the filtering approximation which we draw from
in step 3 (integrating over step 2). Furthermore, we have that in
Algorithm~\ref{alg:asir}, $\omega_{t|t+1}^k = \omega_{t|t+1}(x_t^k)\pi_t^k $ and
$\omega_{t+1}^k = \omega_{t+1}(x_{t+1}^k,x_t^k) $.

\begin{lemma} \label{lemma:asir}%
\[
E[\widehat{p}^{A}(y_{t}|y_{1:t-1})|{\cal A}_{t-1}]=\sum%
\limits_{k=1}^{M}p(y_{t}|x_{t-1}^{k})\pi _{t-1}^{k},
\]%
where the swarm of particles at time $t$ is  ${\cal A}_{t}=\{x_{t}^{k};\pi
_{t}^{k}\}.$
\end{lemma}

\bigskip

\begin{proof}
\begin{eqnarray*}
&&E[\widehat{p}^{A}(y_{t}|y_{1:t-1})\text{ }|\text{ }{\cal A}_{t-1}] \\
&=&E\left[ \sum\limits_{k=1}^{M}\frac{\omega _{t}(x_{t}^{k};\widetilde{x}%
_{t-1}^{k})}{M}\text{ }|\text{ }{\cal A}_{t-1}\right] \left\{
\sum\limits_{j=1}^{M}\omega _{t-1|t}^{j}\right\} \\
&=&\int \omega _{t}(x_{t};\widetilde{x}_{t-1})g(x_{t}|\widetilde{x}%
_{t-1};y_{t})\widehat{g}_{M}^{A}(\widetilde{x}_{t-1}|y_{1:t})dx_{t}d%
\widetilde{x}_{t-1}\left\{ \sum\limits_{j=1}^{M}\omega _{t-1|t}^{j}\right\}
\\
&=&\int \sum\limits_{k=1}^{M}\omega
_{t}(x_{t};x_{t-1}^{k})g(x_{t}|x_{t-1}^{k};y_{t})\frac{\omega
_{t-1|t}(x_{t-1}^{k})}{(\sum\nolimits_{j=1}^{M}\omega _{t-1|t}(x_{t-1}^{j}))%
}dx_{t}\left\{ \sum\limits_{j=1}^{M}\omega _{t-1|t}^{j}\right\} \\
&=&\int \sum\limits_{k=1}^{M}\omega
_{t}(x_{t};x_{t-1}^{k})g(x_{t}|x_{t-1}^{k};y_{t})\omega
_{t-1|t}(x_{t-1}^{k})dx_{t} \\
&=&\sum\limits_{k=1}^{M}\int \frac{p(y_{t}|x_{t})p(x_{t}|x%
_{t-1}^{k})}{g(y_{t}|x_{t}^k)g(x_{t}|x_{t-1}^k, y_t)}
g(x_{t}|x_{t-1}^{k},y_{t})g(y_{t}|x_{t-1}^{k})\pi
_{t-1}^{k}dx_{t},
\end{eqnarray*}%
so%
\begin{eqnarray*}
E[\widehat{p}^{A}(y_{t}|y_{1:t-1})\text{ }|\text{ }{\cal A}_{t-1}]
&=&\sum\limits_{k=1}^{M}\pi _{t-1}^{k}
\int p(y_{t}|x_{t})p(x_{t}|x_{t-1}^{k})dx_{t} \\
&=&\sum\limits_{k=1}^{M}p(y_{t}|x_{t-1}^{k})\pi _{t-1}^{k}.
\end{eqnarray*}
\end{proof}

\begin{lemma}
\[
E[\widehat{p}^{A}(y_{t-h:t}|y_{1:t-h-1})|{\cal A}_{t-h-1}]=\sum%
\limits_{k=1}^{M}p(y_{t-h:t}|x_{t-h-1}^{k})\pi _{t-h-1}^{k}
\]
\end{lemma}

\begin{proof}
(by induction)

{\bf A. Show true for case} ${\bf h=1},$
\begin{eqnarray*}
E[\widehat{p}^{A}(y_{t-1:t}|y_{1:t-2})|{\cal A}_{t-2}] &=&E[\widehat{p}%
^{A}(y_{t}|y_{1:t-1})\widehat{p}^{A}(y_{t-1}|y_{1:t-2})|{\cal A}_{t-2}] \\
&=&E\left[ E[\widehat{p}^{A}(y_{t}|y_{1:t-1})\text{ }|\text{ }{\cal A}_{t-1}]%
\text{ }\widehat{p}^{A}(y_{t-1}|y_{1:t-2})\text{ }|\text{ }{\cal A}_{t-2}%
\right] .
\end{eqnarray*}%
The inner integral is,
\[
E[\widehat{p}^{A}(y_{t}|y_{1:t-1})\text{ }|\text{ }{\cal A}%
_{t-1}]=\sum\limits_{k=1}^{M}p(y_{t}|x_{t-1}^{k})\pi _{t-1}^{k}.\text{,}
\]%
from Lemma \ref{lemma:asir}. Hence,%
\begin{eqnarray*}
&&E[\widehat{p}^{A}(y_{t-1:t}|y_{1:t-2})|{\cal A}_{t-2}] \\
&=&E\left[ \left\{ \tsum\limits_{k=1}^{M}p(y_{t}|x_{t-1}^{k})\pi
_{t-1}^{k}\right\} \left\{ \tsum\limits_{i=1}^{M}\frac{\omega _{t-1}^{i}}{M}%
\right\} \text{ }|\text{ }{\cal A}_{t-2}\right] \left\{
\tsum\limits_{j=1}^{M}\omega _{t-2|t-1}^{j}\right\} \\
&=&E\left[ \left\{ \tsum\limits_{k=1}^{M}p(y_{t}|x_{t-1}^{k})\frac{\omega
_{t-1}^{k}}{\tsum\nolimits_{i=1}^{M}\omega _{t-1}^{i}}\right\} \left\{
\tsum\limits_{i=1}^{M}\frac{\omega _{t-1}^{i}}{M}\right\} \text{ }|\text{ }%
{\cal A}_{t-2}\right] \left\{ \tsum\limits_{j=1}^{M}\omega
_{t-2|t-1}^{j}\right\} \\
&=&E\left[ \left\{ \frac{1}{M}\tsum\limits_{k=1}^{M}p(y_{t}|x_{t-1}^{k})%
\omega _{t-1}^{k}\right\} \text{ }|\text{ }{\cal A}_{t-2}\right] \left\{
\tsum\limits_{j=1}^{M}\omega _{t-2|t-1}^{j}\right\} \\
&=&\left\{ \tsum\limits_{j=1}^{M}\omega _{t-2|t-1}^{j}\right\} \tint
p(y_{t}|x_{t-1})\omega _{t-1}(x_{t-1};\widetilde{x}_{t-2})g(x_{t-1}|%
\widetilde{x}_{t-2};y_{t-1})\widehat{g}_{M}^{A}(\widetilde{x}%
_{t-2}|y_{1:t-1})dx_{t-1}d\widetilde{x}_{t-2} \\
&=&\left\{ \tsum\limits_{j=1}^{M}\omega _{t-2|t-1}^{j}\right\} \int
\tsum\nolimits_{k=1}^{M}p(y_{t}|x_{t-1})\omega
_{t-1}(x_{t-1};x_{t-2}^{k})g(x_{t-1}|x_{t-2}^{k};y_{t-1})\frac{%
g(y_{t-1}|x_{t-2}^{k})\pi _{t-2}^{k}}{\tsum\nolimits_{j=1}^{M}\omega
_{t-2|t-1}^{j}}dx_{t-1} \\
&=&\tsum\nolimits_{k=1}^{M}\pi _{t-2}^{k}\int p(y_{t}|x_{t-1})\omega
_{t-1}(x_{t-1};x_{t-2}^{k})g(y_{t-1}|x_{t-2}^{k})g(x_{t-1}|x_{t-2}^{k};y_{t-1})dx_{t-1}
\\
&=&\tsum\nolimits_{k=1}^{M}\pi _{t-2}^{k}\int
p(y_{t}|x_{t-1})p(y_{t-1}|x_{t-1})p(x_{t-1}|x_{t-2}^{k})dx_{t-1} \\
&=&\tsum\nolimits_{k=1}^{M}p(y_{t-1:t}|x_{t-2}^{k})\pi _{t-2}^{k}\text{ as
required.}
\end{eqnarray*}%
{\bf B Assume that the theorem holds for }$h,$%
\[
E[\widehat{p}^{A}(y_{t-h:t}|y_{1:t-h-1})|{\cal A}_{t-h-1}]=\tsum%
\limits_{k=1}^{M}p(y_{t-h:t}|x_{t-h-1}^{k})\pi _{t-h-1}^{k}
\]%
{\bf C Show that the theorem holds for }$h+1:$

\begin{eqnarray*}
&&E[\widehat{p}^{A}(y_{t-h-1:t}|y_{1:t-h-2})|{\cal A}_{t-h-2}] \\
&=&E\left[ E[\widehat{p}^{A}(y_{t-h:t}|y_{1:t-h-1})\text{ }|\text{ }{\cal A}%
_{t-h-1}]\text{ }\widehat{p}^{A}(y_{t-h-1}|y_{1:t-h-2})\text{ }|\text{ }%
{\cal A}_{t-h-2}\right] \\
&=&E\left[ \tsum\limits_{k=1}^{M}p(y_{t-h:t}|x_{t-h-1}^{k})\pi
_{t-h-1}^{k}\tsum\limits_{i=1}^{M}\frac{\omega _{t-h-1}^{i}}{M}%
\tsum\limits_{j=1}^{M}\omega _{t-h-2|t-h-1}^{j}\text{ }|\text{ }{\cal A}%
_{t-h-2}\right] \text{,}
\end{eqnarray*}%
using Lemma \ref{lemma:asir},
\begin{eqnarray}
&=&E\left[ \left\{ \tsum\limits_{k=1}^{M}p(y_{t-h:t}|x_{t-h-1}^{k})\frac{%
\omega _{t-h-1}^{k}}{\tsum\nolimits_{i=1}^{M}\omega _{t-h-1}^{i}}\right\}
\left\{ \tsum\limits_{i=1}^{M}\frac{\omega _{t-h-1}^{i}}{M}\right\} \text{ }|%
\text{ }{\cal A}_{t-h-2}\right]  \nonumber \\
&&\times \left\{ \tsum\limits_{j=1}^{M}\omega _{t-h-2|t-h-1}^{j}\right\} \\
&=&E\left[ \frac{1}{M}\tsum\limits_{k=1}^{M}p(y_{t-h:t}|x_{t-h-1}^{k})\omega
_{t-h-1}^{k}\text{ }|\text{ }{\cal A}_{t-h-2}\right] \left\{
\tsum\limits_{j=1}^{M}\omega _{t-h-2|t-h-1}^{j}\right\}  \nonumber \\
&=&\left\{ \tsum\limits_{j=1}^{M}\omega _{t-h-2|t-h-1}^{j}\right\} \int
p(y_{t-h:t}|x_{t-h-1})\omega _{t-h-1}(x_{t-h-1};\widetilde{x}_{t-h-2})
\nonumber \\
&&g(x_{t-h-1}|\widetilde{x}_{t-h-2};y_{t-h-1})\widehat{g}_{M}^{A}(\widetilde{%
x}_{t-h-2}|y_{1:t-h-1})dx_{t-h-1}d\widetilde{x}_{t-h-2} \\
&=&\left\{ \tsum\limits_{j=1}^{M}\omega _{t-h-2|t-h-1}^{j}\right\} \int
\tsum\nolimits_{k=1}^{M}p(y_{t-h:t}|x_{t-h-1})\omega
_{t-h-1}(x_{t-h-1};x_{t-h-2}^{k})  \nonumber \\
&&g(x_{t-h-1}|x_{t-h-2}^{k};y_{t-1})\frac{g(y_{t-h-1}|x_{t-h-2}^{k})\pi
_{t-h-2}^{k}}{\tsum\nolimits_{j=1}^{M}\omega _{t-h-2|t-h-1}^{j}}dx_{t-h-1} \\
&=&\tsum\nolimits_{k=1}^{M}\pi _{t-h-2}^{k}\int p(y_{t-h:t}|x_{t-h-1})\omega
_{t-h-1}(x_{t-h-1};x_{t-h-2}^{k})  \nonumber \\
&&g(x_{t-h-1}|x_{t-h-2}^{k};y_{t-1})g(y_{t-h-1}|x_{t-h-2}^{k})dx_{t-h-1},
\end{eqnarray}%
using the definition of $\omega
_{t-h-1}$ (see step 4 of Algorithm),
\begin{eqnarray}
&=&\tsum\nolimits_{k=1}^{M}\pi _{t-h-2}^{k}\int
p(y_{t-h:t}|x_{t-h-1})p(y_{t-h-1}|x_{t-h-1})p(x_{t-h-1}|x_{t-h-2}^{k})dx_{t-h-1}
\nonumber \\
&=&\tsum\nolimits_{k=1}^{M}p(y_{t-h-1:t}|x_{t-h-2}^{k})\pi _{t-h-2}^{k}\text{
as required}  \nonumber
\end{eqnarray}
\end{proof}

\begin{proof}[Proof of \ref{thm: unbiased asir}]
As a consequence we have the lemma that, with $h=t-2$%
\[
E[\widehat{p}^{A}(y_{1:t})|A_{0}]=\tsum\limits_{k=1}^{M}p(y_{1:t}|x_{0}^{k})%
\pi _{0}^{k}
\]%
where $x_{0}^{k}\sim p(x_{0})$ and $\pi _{0}^{k}=1/M$,
\[
E\left[ \tsum\limits_{k=1}^{M}p(y_{1:t}|x_{0}^{k})\pi _{0}^{k}\right] =\tint
p(y_{1:t}|x_{0})p(x_{0})dx_{0}=p(y_{1:t}),
\]%
as required.
\end{proof}


\section{Fully and partially adapted particle filter}\label{app: partially adapted pf}

\subsection{Partially adapted \pf} \label{app: partially adapted pf}
The \papf used in our article is  described in general as follows.
We omit dependence on unknown parameters for clarity.
Suppose that $p(x_{t+1}|x_t)\sim\mathcal{N}(\mu(x_t),\Sigma(x_t))$
and $p(y_{t+1}|x_{t+1})$ is log-concave as a function of $x_{t+1}$.
Let $\ell(x_{t+1}) = \log p(y_{t+1}|x_{t+1})$,
$\lambda(x_{t+1},k) = \ell(x_{t+1}) + \log p(x_{t+1}|x_t^k)$ and let
\begin{align*}
\xbreve_{t+1}^k = \argmax_{x_{t+1}} \lambda(x_{t+1},k),
\quad \Sigmabreve_{t+1}^k = \left ( -
\frac{\partial^2 \lambda(x_{t+1}, k) } {\partial x_{t+1}\partial x_{t+1}}\right)^{-1}_{x_{t+1} = \xbreve_{t+1}^k}
\end{align*}
Then, we take $g(x_{t+1}|\xtilde_t^k,y_{t+1}) = N(x_{t+1}; \xbreve_{t+1}^k, \Sigmabreve_{t+1}^k)$ and
$g(y_{t+1}|x_{t}^k) \propto p(y_{t+1}|\xbreve_{t+1}^k) p(\xbreve_{t+1}^k|x_t^k)  \det(\Sigmabreve_{t+1}^k)^\half $.
This is obtained from the second order approximation
\begin{align*}
\lambda(x_{t+1},k) \doteq \lambda(\xbreve_{t+1},k)+ \log \left ( \det(\Sigmabreve_{t+1}^k)^\half \right )  -\half (x_{t+1} - \xbreve_{t+1}^k)^\prime \left( \Sigmabreve_{t+1}^k\right)^{-1} (x_{t+1} -\xbreve_{t+1}^k) - \log \left ( \det(\Sigmabreve_{t+1}^k)^\half \right )
\end{align*}
The mode $\xbreve_{t+1}^k$ is obtained by  Newton-Raphson iteration with the starting
value $ \mu(x_t^k)$, or some problem specific starting value as in the binomial example.

An alternative iterative scheme to obtain $\xbreve_{t+1}^k $ is based on solving
$\partial \lambda (x)/\partial x =0 =  \partial \ell (x)/\partial x -
\Sigma(x_{t}^k) ^{-1} (x-\mu(x_t^k) ) $. The iteration is given by
\begin{align}\label{eq: itertation alt}
x_{t+1}^k & = \mu(x_{t}^k ) + \Sigma(x_{t}^k)\partial \ell (x_{t+1}^k)/\partial x_{t+1}^k.
\end{align}
A single iteration of \eqref{eq: itertation alt} is usually faster than a single iteration of the Newton-Raphson scheme but the actual speed of convergence depends on the problem. In practice, we can make the iterations to the mode faster by just taking a fixed small number of steps  of either the Newton-Raphson or \eqref{eq: itertation alt}, or by making the convergence criterion less strict. If we take a fixed number of steps then the iterations can be vectorized over $k$.

\subparagraph{Binomial example} For the binomial example discussed in Section~\ref{ss: binomial model} we use
the Newton-Raphson iteration
with starting value $\mu(x_t^k)$ or  $\logit(y_t/m)$ if $m$ is large.

\subsection{Fully adaptive \pf} \label{app: fully adaptive pf}
Full adaptation is possible whenever $p(x_{t+1}|x_t)$ is conjugate in $x_{t+1}$ to
$p(y_{t+1}|x_{t_+1})$.
\subparagraph{Gaussian observation equation}
Suppose that the observation equation is Gaussian with $p(y_{t}|x_t) \sim N(H_tx_t, V_t)$ and the state transition equation is the same as in Section~\ref{app: partially adapted pf}.
Then, from Section~ \ref{app: partially adapted pf}, $\xbreve_{t+1}^k$ and $\Sigmabreve_{t+1}^k$ are obtained explicitly as
\begin{align*}
\Sigmabreve_{t+1}^k & = \left ( H_{t+1}^\prime V_{t+1}^{-1} H_{t+1} + \Sigma(x_t^k)^{-1} \right) ^{-1} , \quad \xbreve_{t+1}^k = \Sigmabreve_{t+1}^k \left ( V_{t+1}^{-1} y_{t+1} + \Sigma(x_t^k)^{-1} \mu(x_t^k) \right ) \ ,
\end{align*}
and $p(y_{t+1}|x_t^k)$ is obtained as in  Section~\ref{app: partially adapted pf}.
\subparagraph{Garch model} We use the notation in Section~\ref{SS: garch with noise}.
It is straightforward to show that
$p(y_{t+1}|x_t,\sigma_t^2) \sim N(0, \sigma_{t+1}^2+\tau^2)$ and that
$p(x_{t+1}|y_{t+1},x_t) \sim N(a_{t+1}, \Delta_{t+1}) $, where
\begin{align*}
\Delta_{t+1}^{-1} & = (\tau^2)^{-1} + (\sigma_{t+1}^2)^{-1} ,  \quad a_{t+1} = \Delta_{t+1} y_{t+1}/\tau^2\ .
\end{align*}

\section{Adaptive sampling schemes} \label{app: adaptive sampling}
This appendix describes the two adaptive sampling schemes used in the paper.
\subsection{Adaptive random walk Metropolis}\label{subsection:ARWM}
The adaptive random walk Metropolis proposal of \cite{roberts:rosenthal:2008} is
\begin{equation} \label{e:arwm proposal}
q_j(\theta;\theta_{j-1}) =\omega_{1j}\phi_d (\theta; \theta_{j-1}, \kappa_1\Sigma_1) + \omega_{2j}\phi_d (\theta; \theta_{j-1}, \kappa_2\Sigma_{2j})
\end{equation}
where $d$ is the dimension of $\theta$ and $\phi_d(\theta; \thetat, \Sigma)$ is a multivariate
$d$ dimensional normal density in $\theta$ with mean $\thetat$ and covariance matrix $\Sigma$.
In \eqref{e:arwm proposal}, $\omega_{1j} = 1$ for $j \leq j_0$, with $j_0$ representing the initial iterations,
$\omega_{1j} = 0.05$ for $j > j_0$ with $\omega_{2j} = 1 -  \omega_{1j}$; $\kappa_1 = 0.1^2/d, \kappa_2 = 2.38^2/d, \Sigma_1 $ is a constant covariance matrix, which is taken as the identity matrix by \cite{roberts:rosenthal:2008} but can be based on the Laplace approximation or some other estimate. The matrix $\Sigma_{2j}$ is the sample covariance matrix of the first $j-1$ iterates. The scalar $\kappa_1$ is meant to achieve a high acceptance rate by moving the sampler locally, while the scalar $\kappa_2$ is considered to be optimal \citep{roberts:gelman:gilks:1997} for a random walk proposal when the target is a multivariate normal. We note that the acceptance probability \eqref{e:adaptive accep prob pf}  for the adaptive random walk Metropolis simplifies to
\begin{equation} \label{e:arwm accep prob}
\alpha(\theta_{j-1},u_{j-1};\theta_{j}^p,u^{p}) = \min \biggl \{1,\frac{p(y|\theta_{j}^p,u_j^p)p(\theta^p)}
{p(y|\theta_{j-1},u_{j-1})p(\theta_{j-1})}\biggr \} \ .
\end{equation}

\subsection{ A proposal density based on a mixture of normals}\label{subsection:AIMH:MN}
The proposal density of the adaptive independent Metropolis-Hastings approach of
\citet{giordani:kohn:2010} is a mixture with four terms of the form
\begin{equation}\label{eq:aimh proposal}
q_j(\theta)=\sum_{k=1}^4\omega_{kj}g_{k}(\theta| \lambda_{kj} )\,
\quad  \quad \omega_{kj} \ge 0, \quad\text{for} \quad k=1, \dots, 4
\quad \text{and}\quad  \sum_{k=1}^4 \omega_{kj} = 1\ ,
\end{equation}
with $\lambda_{kj}$ the  parameter vector for the density $g_{kj}(\theta;\lambda_{kj})$.
The sampling scheme is run in two stages, which are described below. Throughout each stage, the parameters in the
first two terms are kept fixed.  The first term $g_{1}(\theta| \lambda_{1j})$ is an estimate of the target density and
the second term $g_{2}(\theta| \lambda_{2j})$ is a heavy tailed version of $g_{1}(\theta| \lambda_{1j})$.
The third term $g_{3}(\theta| \lambda_{3j})$ is an estimate of the target that is updated or adapted as the simulation
progresses and the fourth term $g_{4}(\theta| \lambda_{4j})$ is a heavy tailed version of the third term.
In the first stage $g_{1j}(\theta; \lambda_{1j})$ is a Gaussian density constructed from a preliminary run,
of the three component adaptive random walk. Throughout, $g_{2}(\theta| \lambda_{2j})$ has
the same component means and probabilities as $g_{1}(\theta| \lambda_{1j})$, but its component covariance matrices are ten times those of $g_{1}(\theta| \lambda_{1j})$. The term $g_{3}(\theta| \lambda_{3j})$ is a mixture of normals
and $g_{4}(\theta| \lambda_{4j})$ is also a mixture of normals obtained by taking its component probabilities and means equal to those of $g_{3}(\theta| \lambda_{3j})$, and its component covariance matrices equal to 20 times those of $g_{3}(\theta| \lambda_{3j})$. The first stage begins by using $g_{1}(\theta|\lambda_{1j})$ and $g_{2}(\theta| \lambda_{2j})$  only with,  for example, $\omega_{1j} = 0.8$ and $\omega_{2j} = 0.2$, until there is a sufficiently large number of iterates to form $g_{3}(\theta| \lambda_{3j})$. After that we set $\omega_{1j} = 0.15, \omega_{2j} = 0.05, \omega_{3j} = 0.7 $ and $\omega_{4j} = 0.1$. We begin with a single normal density for $g_{3}(\theta| \lambda_{3j})$ and as the simulation progresses we add more components up to a maximum of six according to a schedule that depends on the ratio of the number of accepted draws to the dimension of $\theta$.

In the second stage, $g_{1}(\theta| \lambda_{1j})$ is set to the value of $g_{3}(\theta| \lambda_{3j})$ at the end of the first stage and $g_{2}(\theta| \lambda_{2j})$ and $g_{4}(\theta| \lambda_{4j})$ are constructed as described above. The heavy-tailed densities $g_{2}(\theta| \lambda_{2j})$ and $g_{4}(\theta| \lambda_{4j})$ are included as a
defensive strategy to get out of  local modes and to explore the sample space of the target distribution more effectively.

It is computationally too expensive to update $g_{3}(\theta| \lambda_{3j})$ (and hence $g_{4}(\theta| \lambda_{4j})$) at every iteration so we update them according to a schedule that depends on the problem and the size of the parameter vector.

\subsection{Proof of the convergence of the \aimh sampling scheme} \label{ss: convergence}
The following convergence results hold for the \aimh{}
sampling scheme described in Appendix~\ref{subsection:AIMH:MN} (and more fully in
\citet{giordani:kohn:2010})
when it is combined with the ASIR \pf. They  follow from Theorems~1 and 2 of
\citet{giordani:kohn:2010}.
Let $\Theta$ be the parameter space of $\theta$.
\begin{thm} \label{thm: convergence}
Suppose that there exists a constant $0<C<\infty$ that does not depend on $t=1,\dots, T,
\theta\in \Theta$ and the number of iterates $j$ such that
\begin{align}
g(y_{t+1}|x_t;\theta) & \leq C ,  \label{eq: gyx ineq} \\
\frac{p(y_{t+1}|x_{t+1}; \theta)p(x_{t+1}|x_t;\theta)}
{g (y_{t+1}|x_{t};\theta)g(x_{t+1}|y_{t+1},x_t;\theta) } & \leq C ,
\label{eq: y over g ineq} \\
p(\theta)/ q_j(\theta) & \leq C \ . \label{eq: p over q ineq}
\end{align}
Then,
\begin{enumerate}
\item
The simulated likelihood is bounded uniformly in $\theta \in \Theta$.
\item
The iterates $\theta_j$ of the \aimh{} sampling scheme converge to a sample from $p(\theta|y)$ in the sense that
\begin{align} \label{e:theorem1}
\sup_{A \subset \Theta} \mid \Pr(\theta_j  \in A) - \int_A p(\theta \mid y )d\theta  \mid & \ra 0 \quad \text{as} \quad j \ra \infty.
\end{align}
for all measurable sets $A$ of $\Theta$.
\item
Suppose that $h(\theta)$ is a measurable function of $\theta$
that is square integrable with respect to the density $g_2$. Then, almost surely,
\begin{align} \label{e:theorem2}
\frac{1}{n} \sum_{j=1}^n h (\theta_j) \ra \int h(\theta)p(\theta|y) d\theta  \quad  \text{as   } \quad n \ra
\infty.
\end{align}
\end{enumerate}
\begin{proof}
\begin{align*}
p_S (y|\theta,u) & = \prod_{t=0}^{T-1} p_S(y_{t+1}|y_{1:t};\theta,u)
 \leq C^{2T} \quad \text{because}\quad
p_S (y_t|y_{1:t-1};\theta)  \leq C^2
\end{align*}
from \eqref{asir_predlik_est} and our assumptions.
This shows that the simulated  likelihood $p_S (y|\theta,u)$ is bounded
and the result now follows from \citet{giordani:kohn:2010}.
\end{proof}
\end{thm}

We note that as in
\cite{giordani:kohn:2010} it is straightforward to choose the proposal density $q_j(\theta)$ as a mixture with one component that is at least as heavy tailed as $p(\theta)$ to ensure that \eqref{eq: p over q ineq} holds.

The next corollary gives a condition for equations \eqref{eq: gyx ineq}
and \ref{eq: y over g ineq} to hold for the \spf and the  \fapf.
\begin{corollary}\label{cor: convergence}
Suppose that for all $y_t, x_t, t=1, \dots, T$ and $\theta \in \Theta$, there exists  a constant   $C_1 > 0$ such that
\begin{align} \label{eqn: p y given x}
p(y_t|x_t;\theta)  \leq C_1.
\end{align}
Then equations \eqref{eq: gyx ineq} and
\eqref{eq: y over g ineq} hold for the \spf and the \fapf.
\begin{proof}
We have
\begin{align*}
p(y_{t+1}|x_t;\theta) & = \int p(y_{t+1}|x_{t+1};\theta)p(x_{t+1}|x_t;\theta) dx_{t+1} \leq C_1
\end{align*}
and the result follows for the \spf and the \fapf.
\end{proof}
\end{corollary}

We note that usually $p(y_t|x_t; \theta)$ is uniformly bounded in $y_t, x_t$ and $\theta$ for $t=1, \dots, T$.
 This is true for the models in Sections~\ref{S: comparing SIR with ASIR} and
\ref{section:algorithm:comparison}.

We now construct a \papf that satisfies equations \eqref{eq: gyx ineq} and
\eqref{eq: y over g ineq}. Suppose that $g_0(y_{t+1}|x_t;\theta)$ and
$g_0(x_{t+1}|y_{t+1},x_t;\theta)$ correspond to a \papf
which we refer to as $g_0$, e.g. the \papf described in Section
\ref{app: partially adapted pf}. Let $0 < \epsilon < 1 $. Now construct the \papf $g$ as a mixture taking the value $g_0$ with probability $1-\epsilon$ and  being the \spf with probability $\epsilon$. That is,
\begin{align}\label{eqn: papf mixture}
g(y_{t+1}|x_t;\theta ) g(x_{t+1}|x_t,y_{t+1};\theta)& = \epsilon p(x_{t+1}|x_t) + (1-\epsilon) g_0(y_{t+1}|x_t) g_0(x_{t+1}|x_t,y_{t+1};\theta)  \ .
\end{align}

\begin{corollary}\label{cor: condns for papf}
Suppose  equation \eqref{eqn: p y given x} holds and the \papf is defined by
equation \eqref{eqn: papf mixture}.
Then, equations \eqref{eq: gyx ineq} and \eqref{eq: y over g ineq} hold.

The proof is straightforward.
\end{corollary}

Usually, we would  take $\epsilon $ quite small so that most of the time the \papf $g_0$ is used. Using the mixture \papf ensures that the simulated likelihood is bounded which is important to successfully use the \aimh to sample the parameters.

\section{Marginal likelihood evaluation using bridge and importance sampling} \label{A: BS and IS}
Suppose that $q(\theta)$ is an approximation to $p(\theta| y )$ which can be evaluated explicitly. Bridge sampling \citep{meng:wong:1996} estimates the marginal likelihood as follows. Let
\begin{equation}\notag
t(\theta)=\left(\dfrac{p(y|\theta)p(\theta)}{U}+q(\theta)\right)^{-1},
\end{equation}
where $U$ is a positive constant. Let
\begin{align} \label{eq: bridge1}
\begin{split}
A & = \int t(\theta)q(\theta)p(\theta\mid y)d\theta \ . \qquad \text{Then,} \\
A &  = \frac{A_1}{p(y)} \qquad \text{where} \qquad A_1  = \int t(\theta)q(\theta)p(y\mid \theta) p(\theta) d\theta \ .
\end{split}
\end{align}
Suppose the sequence of iterates $\{\theta^{(j)},j=1,\ldots,M\}$ is generated from the posterior density $p(\theta|y)$ and a second sequence of iterates $\{\ttheta^{(k)},k=1,\ldots,M\}$ is generated from $q(\theta)$. Then
\begin{align}\notag
\widehat{A} & = \dfrac{1}{M}\sum_{j=1}^{M}t(\theta^{(j)})q(\theta^{(j)}), \quad
\widehat{A}_1  =\dfrac{1}{M}\sum_{k=1}^{K} t(\theta^{(k)})p(y|\theta^{(k)})p(\theta^{(k)})\quad \text{and}\quad
\widehat{p}_{BS}(y)=\dfrac{\widehat{A}_1}{\widehat{A}}
\end{align}
are estimates of $A$ and $A_1$ and $\widehat{p}_{BS}(y)$ is the bridge sampling estimator of the marginal likelihood $p(y)$.

In adaptive sampling, $q(\theta)$ is  the mixture of normals proposal.
Although $U$ can be any positive constant, it is more efficient if $U$ is a reasonable estimate of $p(y)$. One way to do so is to take
$
\Uhat  = p(y|\theta^*)p(\theta^*)/q(\theta^*)$,
where $\theta^*$ is the posterior mean of $\theta$ obtained from the posterior simulation.

An alternative method to estimate of the marginal likelihood $p(y)$ is to use importance sampling based on the proposal distribution $q(\theta)$ \citep{geweke:1989,chen:shao:1997}. That is,
\begin{equation}\nonumber
\widehat{p}_{IS}(y)=\dfrac{1}{K}\sum_{k=1}^{K}\dfrac{p(y|\theta^{(k)})p(\theta^{(k)})}{q(\theta^{(k)})}.
\end{equation}
Since our proposal distributions have at least one heavy tailed component, the importance sampling ratios are likely to be bounded and well-behaved, as in the examples in this paper.

\section{Implementation details} \label{app: Implementation}
We coded most of the algorithms in MATLAB, with a small proportion of the code
written using C/Mex files. We carried out the estimation  on an SGI cluster with 42 compute nodes. Each of them is an SGI Altix XE320 with two Intel Xeon X5472
(quad core 3.0GHz) CPUs with at least 16GB memory. We ran parallel jobs using up to eight processors and MATLAB 2009.

\end{appendices}

\end{document}